\documentclass[11pt,a4paper]{article}

\usepackage{jheppub}

\def\w{{\omega}}

\def\a{{\alpha}}

\def\s{\sqrt}

\def\CO{{\cal O}}

\def\CN{{\cal N}}

\def\pp{\partial}

\def\be{\begin{equation}}
\def\ee{\end{equation}}
\def\ba{\begin{eqnarray}}
\def\ea{\end{eqnarray}}
\def\bal#1\eal{\begin{align}#1\end{align}}

\def\dg{\dagger}
\def\db{\bar{\partial}}

\def\r{\rightarrow}

\def\LR{\Leftrightarrow}
\def\f {\frac}

\def\ddd{\cdot\cdot\cdot}
\def\no{\nonumber \\}

\def\la{\langle}
\def\lb{\rangle}

\def\r{\rightarrow}

\def\lr{\leftrightarrow}

\def\q{\quad}
\def\qq{\quad\quad}

\def\z{\bar{z}}

\begin{document}
%\begin{titlepage}
%\thispagestyle{empty}

\begin{flushright}
UT-Komaba17-2
\\
YITP-17-31
\\
IPMU17-0047
\\
\end{flushright}

%\bigskip

%\begin{center}
%{{\textbf{\large Bulk local states on the black holes in AdS/CFT}}}\\
%\vspace{2cm}
%Kanato Goto$^{a}$ and Tadashi Takayanagi$^{b,c}$
%\vspace{1cm}
\title{CFT descriptions of bulk local states \\ %construction
in the AdS black holes}
\author{Kanato Goto$^a$ and}
\author{Tadashi Takayanagi$^{b,c}$}
%{\it
%$^{a}$ Institute of Physics, The University of Tokyo\\ Komaba, Meguro-ku, Tokyo 153-8902, Japan\\
%$^{b}$Yukawa Institute for Theoretical Physics,\\
%Kyoto University, Kyoto 606-8502, Japan\\
%$^{c}$Kavli Institute for the Physics and Mathematics of the Universe,\\
%University of Tokyo, Kashiwa, Chiba 277-8582, Japan\\
%}
\affiliation[a]{Institute of Physics, The University of Tokyo Komaba, Meguro-ku, Tokyo 153-8902, Japan}
\affiliation[b]{Center for Gravitational Physics, Yukawa Institute for Theoretical Physics (YITP),
Kyoto University, Kyoto 606-8502, Japan}
\affiliation[c]{Kavli Institute for the Physics and Mathematics of the Universe,
University of Tokyo, Kashiwa, Chiba 277-8582, Japan}
\abstract{We present a new method for reconstructing CFT duals of states excited by the bulk local operators in the three dimensional AdS black holes in the AdS/CFT context. As an important procedure 
for this, we introduce  a map between the bulk points in AdS and those on the boundary where CFT lives. This gives a systematic and universal way to express bulk local states even inside black hole interiors. Our construction allows us to probe the interior structures of black holes purely from the CFT calculations. We analyze bulk local states in the single-sided black holes as well as the double-sided black holes.}

\maketitle
\flushbottom

%\addtocontents{toc}{\protect\enlargethispage{2\baselineskip}}

\newpage

%\setcounter{page}{1}
%\pagenumbering{arabic}
\section{Introduction}
AdS/CFT correspondence gives a UV complete description of the quantum gravity in the $d+1$-dimensional asymptotically AdS spacetime in terms of the large $c$ conformal field theories defined on its $d$-dimensional boundary \cite{Mal}.
From the perspective of the AdS/CFT, the notion of the bulk spacetime is emergent from the Hilbert space of the dual CFT in the large $c$ limit. It is still not well understood how CFTs can describe classical spacetimes. 
Though various attempts have been made to probe the bulk geometry from CFT side, the most direct approach to this is to consider a CFT dual of a local field in the bulk which is most fundamental element of the low energy effective theory of quantum gravity.

Especially in the last few years, motivated by firewall arguments \cite{AMPS}\cite{AMPSS}\cite{MP0}, it has been actively debated whether CFTs can contain operators dual to the local fields in the AdS black hole spacetime, particularly behind the horizon.

There is a well known method to reconstruct bulk local operators so called Hamilton-Kabat-Lifschytz-Lowe(HKLL) construction \cite{HKLL0}\cite{HKLL1}\cite{HKLL2}\cite{HKLL3}. They construct bulk local operators as  CFT operators smeared over boundary regions which are set by the bulk positions. However, one has to solve the bulk equation of motion to give an expression for the bulk local operator, thus it is not satisfactory for understanding the emergence of the bulk spacetime. Furthermore, its expression diverges when one attempts to construct a bulk local operator in the black hole (or Rindler) spacetime. To avoid its divergence, they smeared CFT operator along the complexified spatial directions, but this is not a convenient prescription in ordinary quantum field theories.

Their prescription of reconstructing black hole interiors basically requires the use of smearing functions with support on both of the left and the right boundaries in the Kruskal diagram\footnote{There are some attempts to apply it for the single-sided black holes formed by null-shell collapse, for example see \cite{HKLL2}\cite{RS}.}. Physically meaningful description of the fully Kruskal extended spacetime can be obtained by two copies of the CFTs \cite{Ete}, thus it is not clear whether the Hilbert space of a single CFT can describe black hole interior. 

An important proposal for reconstructing the bulk local operators inside the horizons of single-sided black holes was made by Papadodimas and Raju \cite{PR1}\cite{PR2}\cite{PR3}. Their proposal is free from divergences since they work in the momentum space instead of expressing bulk local fields as integrals of primary operators in the position space. They gave a recipe to reconstruct bulk local fields inside the black hole horizons on generic thermal states in CFT whose techniques have a connection with Tomita-Takesaki theory of operator algebra. Their construction implies the state-dependent map between the Hilbert space of CFT and AdS whose physical validity is controversial and has been argued in \cite{H}\cite{MP} for instance.

In this paper, we present another method to reconstruct the bulk local operator $\phi_\a(r,\vec{x})$ (or equivalently states excited by $\phi_\a(r,\vec{x})$ which we call bulk local states) in the three dimensional BTZ black hole background. This is an application of the construction method in the pure AdS spacetime which is proposed recently in \cite{cMERA}\cite{NO} (see also \cite{Wan}\cite{GMT})\footnote{There is also a slightly different proposal \cite{Ver} which uses the full Virasoro symmetry to construct bulk local states. They discussed the gravitational dressing of the bulk local states using their construction \cite{VLT}.}\footnote{These arguments mainly focus on the leading order of the $1/c$ expansions of CFTs. For the $1/c$ corrections of bulk locals states, see \cite{NO2}\cite{VLT}. }.

A key feature of our method is to construct the bulk local states via the map between points in the bulk $(r,\vec{x})$ and those on the Euclidean boundary as we will explain it later. This map is obtained from the geodesics which connect between boundary points and bulk ones\footnote{The same map appears in the context of emergent hyperbolic space from an optimization of Euclidean path-integral \cite{OPT}, which was recently conjectured to be a new explanation of AdS/CFT. Our results in this paper support this conjecture.}. We do not need to solve bulk equations of motion for representing bulk local fields; one can obtain their expressions just by summing up CFT operators inserted at the boundary points which correspond to the bulk positions $(r,\vec{x})$ through the map. Our method enables to systematically reconstruct bulk local states from various CFT states dual to black hole spacetimes. Especially it allows us to probe the black hole interiors.

This paper is organized as follows. In section \ref{sec1}, we discuss the construction of bulk local states in the pure AdS spacetime using the global conformal symmetry. First we review the construction in the global coordinate and introduce the map between the bulk $(r,\vec{x})$ and boundary $(z_0,\z_0)$ we explained above. Finally, we give a local expression around $(z_0,\z_0)$ for the bulk local states at $(r,\vec{x})$ using the map. In \ref{Rindler}, we consider to construct the bulk local operators on the thermofield double state which is dual to the Rindler coordinate. We apply the same method to the bulk local states in the Rindler coordinate by using the map between the bulk  and the boundary coordinates in the Rindler spacetime. In the section \ref{BTZ}, we attempt to apply our strategy to the black hole states. In \ref{TFD}, we first consider the eternal black holes dual to the thermofield double states which are entangled states of two copies of the CFT. The construction of bulk local states in these black holes provides us a CFT perspective of the black hole singularity. In \ref{Bou}, we apply our method to the single-sided black holes dual to the CFT boundary states which is made by $\mathbb{Z}_2$ identifications of the double-sided black holes. Finally in  \ref{Hea}, we analyze the bulk local operators on the single-sided black holes created by heavy primary operators $\CO_H$ in CFT. To evaluate two point functions, we approximate the correlator by the vacuum conformal block in the channel $\CO_H\CO_H\r\CO_\a\CO_\a$ in the semiclassical  limit \cite{Hart}\cite{FKW1}\cite{FKW2}. Our prescription is valid as far as this semi-classical approximation holds. Differently from the double-sided black holes, we find a implication of ``phase transition'' where our semi-classical prescription will break down if one attempts to move bulk local operators %to the ``second asymptotic region'' .
deep in the bulk direction. In appendix A, we present the details on the algebraic aspects of bulk local states. In particular we construct the bulk local state in Poincare AdS and show that it agrees with the HKLL prescription. In appendix B, we give a comparison with the mode expansion expression of
the bulk local operator.

We believe that our prescription presents a useful tool for probing the interior structures of single-sided black holes which have not been fully analyzed yet. While they depend on the details of CFTs, the map we used for constructing bulk local states directly connects the physics in the bulk interior and physics on the boundary. This enables us to analyze several aspects of the black hole interior purely from the CFT calculations.

\newpage
\section{Bulk local states on the pure AdS}\label{sec1}
A bulk local operator $\hat{\phi}^{\rm CFT}_\a(\boldsymbol{x})$ on a background geometry $g_{\mu\nu}$ is an operator in the CFT which is dual to a local field $\phi_\a(\boldsymbol{x})$ on the geometry $g_{\mu\nu}$ of the AdS spacetime. In this paper, we assume that the local operator is a scalar and write bulk coordinates as $\boldsymbol{x}=(r,\vec{x})$ where $r$ is the radial coordinate and $\vec{x}=(x^\mu)_{\mu=0,\ddd,d-1}$ parameterizes boundary directions. \\
We mainly focus on the leading order of $1/c$ expansion in the large $c$ CFTs where fields in AdS become free fields. Bulk local operators should obey the free scalar field equations of motions on the background geometry $g_{\mu\nu}$
\ba
(\Box_{g_{\mu\nu}}+m^2_\a)\hat{\phi}^{\rm CFT}_\a(\boldsymbol{x})=0.\label{eq}
\ea
 A bulk local state which we mainly discuss in this paper is a state excited by $\hat{\phi}^{\rm CFT}_\a(\boldsymbol{x})$ from a CFT state $|\Psi_{g_{\mu\nu}}\lb$ which is dual to a geometry: $g_{\mu\nu}$ of  AdS.
In this section, we focus on the pure AdS spacetime.
 Especially in subsection \ref{Global}, we will work on the global coordinate which is dual to the vacuum state in the CFT defined on the the boundary of the global AdS: $|\Psi_{{\rm global}}\lb=|0\lb_{\rm CFT}$. In the global coordinate, a bulk local state is expressed as $|\phi_\a(\boldsymbol{x})\lb_{\rm global}\equiv \hat{\phi}^{\rm CFT}_\a(\boldsymbol{x})|0\lb_{\rm CFT}$ where $\boldsymbol{x}$ denotes a bulk position on the global coordinate. In subsection \ref{Rindler}, we discuss the Rindler coordinate which is dual to the thermofield double state: $|\Psi_{{\rm Rindler}}\lb=|\Psi_{\rm TFD}\lb$. The bulk local state in the Rindler coordinate is expressed as $|\phi_\a(\boldsymbol{x})\lb_{\rm Rindler}\equiv \hat{\phi}^{\rm CFT}_\a(\boldsymbol{x})|\Psi_{\rm TFD}\lb$ where $\boldsymbol{x}$ is a bulk point on the Rindler coordinate. In the section \ref{BTZ}, we attempt to construct bulk local states $|\phi_\a(\boldsymbol{x})\lb_{\rm BH}\equiv \hat{\phi}^{\rm CFT}_\a(\boldsymbol{x})|\Psi_{\rm BH}\lb$ from CFT states $|\Psi_{\rm BH}\lb$ dual to the black hole spacetimes.

In the dictionary of AdS/CFT, a local scalar field with mass $m_\a$ in AdS corresponds to a spin-less primary field in CFT with conformal dimension $\Delta_\a$ when the local field in AdS approaches the boundary,
\bal
\phi_\a(\boldsymbol{x})\LR \CO_\a(\vec{x})\qq
m^2_\a R^2=\Delta_\a(\Delta_\a-d),\label{corre}
\eal
where $\a$ denotes the label of the primaries. This correspondence implies that we can reconstruct bulk local fields from the corresponding primary fields in the CFT. There is a well-known method to reconstruct bulk local fields, so called ``HKLL" construction, which expresses bulk local fields as smeared primary operators \cite{HKLL0}\cite{HKLL1}\cite{HKLL2}. In this paper, we use another method proposed in \cite{cMERA}\cite{NO} (see also \cite{Wan}\cite{Ver}\cite{GMT}) 
instead of using HKLL method. It makes use of the correspondence of the symmetry of AdS and CFT;  isometry of AdS and conformal symmetry. We review this construction method briefly.

First consider scalar fields at the center of the AdS $\hat{\phi}_\a(0)=e^{-i\boldsymbol{P}\cdot\boldsymbol{x}}\hat{\phi}_\a(\boldsymbol{x})e^{i\boldsymbol{P}\cdot\boldsymbol{x}}$. The generators of the conformal symmetry are organized as $(D,M_{\mu\nu},P_\mu,K_\mu)$. By using the correspondence with the AdS isometry, we can find a subgroup of the symmetry which keeps the center of the AdS invariant. We can easily see that it is generated by $M_{\mu\nu}$ and $P_{\mu}+K_{\mu}$  \cite{cMERA}\cite{NO}. Thus the dual of a scalar field in AdS: $\hat{\phi}^{\rm CFT}_\a(0)$ should satisfy
\bal
[M_{\mu\nu},\hat{\phi}^{\rm CFT}_\a(0)]=[P_{\mu}+K_{\mu},\hat{\phi}^{\rm CFT}_\a(0)]=0.\label{condi}
\eal
In the following subsections, we will see explicit forms of the solutions  of (\ref{condi}) which satisfy the correspondence (\ref{corre}).
\subsection{Global AdS}\label{Global}
First we review the construction of the bulk local states in the global coordinate \cite{cMERA}\cite{NO}. For simplicity we focus on AdS$_3$/CFT$_2$ while we can also do the same argument in higher dimensions. We write the coordinate of global AdS$_3$ as $\boldsymbol{x}=(\rho,\vec{x})$ where $\vec{x}=(\phi,t)$ is the boundary coordinate of global AdS$_3$.The metric of the global coordinate can be written as
\bal
ds^{2}=&R^{2}(-\cosh^{2}\rho dt^{2}+d\rho^{2}+\sinh^{2}\rho
d\phi^{2})\no
&0<\rho<\infty,0<\phi<2\pi,-\infty<t<\infty.
\eal
The global AdS$_3$ spacetime has an isometry $SO(2,2)=SL(2,\mathbb{R})\times SL(2,\mathbb{R})$ which is generated by $(L_{-1},L_{0},L_{1})$ and $(\bar{L}_{-1},\bar{L}_{0},\bar{L}_{1})$. These are explicitly given by the following Killing vectors:
\bal
L_0&=i\pp_+\no
L_{\pm1}&=ie^{\pm ix^{+}}\biggl[\f{\cosh2\rho}{\sinh2\rho}\pp_+-\f{1}{\sinh2\rho}\pp_-\mp\f{i}{2}\pp_\rho\biggl]\no
\bar{L}_0&=i\pp_-\no
\bar{L}_{+1}&=ie^{\pm ix^{-}}\biggl[\f{\cosh2\rho}{\sinh2\rho}\pp_--\f{1}{\sinh2\rho}\pp_+\mp\f{i}{2}\pp_\rho\biggl]\label{Kill},
\eal
where $x^{\pm}=t\pm \phi$. %and we expressed the generators in the elliptic basis (see also Appendix A).
We wick-rotate the boundary coordinate $x^{\pm}=-i\tau\pm \phi$ and introduce the coordinate $(z,\z)$ on the Euclidean plane defined on boundary of global AdS$_{3}$ as
\ba
z=e^{ix^{+}}=e^{\tau+i\phi},\ \z=e^{ix^{-}}=e^{\tau-i\phi}.
\ea
 Coordinate $(z,\z)$ cuts the Euclidean plane as $\mathbb{R}\times \mathbb{S}$ and states in the CFT are defined by the path-integral from the origin to the unit circle which corresponds to a time slice $\tau=0$. Especially in AdS/CFT, the vacuum state $|0\lb_{\rm CFT}$ defined by the path-integral along the Euclidean time $\tau$ without any operator insertions corresponds to the  global AdS spacetime with no excitations.
From (\ref{Kill}), we can confirm that AdS$_{3}$ isometry is dual to the global part of the $2d$ conformal symmetry whose generators are written as
\ba
L_n=\oint _{\mathcal{C}}\f{dz}{2\pi i}z^{n+1}T(z),\q
\bar{L}_n=\oint _{\mathcal{C}}\f{d\z}{2\pi i}\z^{n+1}\bar{T}(\z)\label{conformalgenerator}
\ea
where $\mathcal{C}$ is a unit circle and the global part corresponds to $n=-1,0,1$.

Now we consider a bulk local state $|\phi_\a(\boldsymbol{x})\lb=\hat{\phi}^{\rm CFT}_\a(\boldsymbol{x})|0\lb_{\rm CFT}$ which is dual to a global AdS spacetime locally excited by a local field in AdS: $\phi_\a(\boldsymbol{x})$. In the following sections, we often write $\hat{\phi}^{\rm CFT}_{\a}|0\lb_{\rm CFT}$ just as $\hat{\phi}_{\a}|0\lb$ unless it causes any confusion. First we work on the center of AdS: $\boldsymbol{x}=(\rho,\phi,t)=0$.
The conditions for the bulk local operators (\ref{condi}) can be rewritten with generators $(L_{-1},L_{0},L_{1})$ and $(\bar{L}_{-1},\bar{L}_{0},\bar{L}_{1})$ as
\ba
(L_0-\bar{L}_0)|\phi_\a\lb=(L_{\pm 1}+\bar{L}_{\mp 1})|\phi_\a\lb=0 .\label{Ishibshii}
\ea
The solutions for these equations can be constructed from primary states $|\CO_\a\lb=\CO_\a(0)|0\lb$ dual to the bulk local operators $\hat{\phi}_{\a}$ and its descendants. %\footnote{Primary states can be obtained by inserting primary operators at the origin of Euclidean boundary; $(z,\z)=0$ i.e, $\tau=-\infty$ in the euclidean time of the global coordinate.}.
They are explicitly written as Ishibashi states with respect to $SL(2,\mathbb{R})$ algebra with $(-1)^{k}$ twist.
\ba
|\phi_\a\lb=\sum_{k=0}^{\infty}(-1)^k\f{\Gamma(\Delta)}{k!\Gamma(k+\Delta)}%(-1)^k
L_{-1}^{k}\bar{L}_{-1}^{k}|\CO_{\a}\lb.\label{Ishibashi}
\ea
\subsection{$SL(2,\mathbb{R})$ transformations to different bulk points}
Bulk local states at different bulk points are related by the $SL(2,\mathbb{R})$ transformations $g(\rho,\phi,t)$. From the correspondence between Killing vectors in AdS (\ref{Kill}) and the conformal symmetry generators in CFT (\ref{conformalgenerator}), we can obtain the expression for a bulk local state at a bulk point $(\rho,\phi,t)$:
\bal
|\phi_\a(\rho,\phi,t)\lb&=e^{i(L_0+\bar{L}_0)t}e^{i(L_0-\bar{L}_0)\phi}e^{\mathchar`-
\rho\frac{L_{1}-\bar{L}_{\mathchar`-1}+\bar{L}_{1}-L_{\mathchar`-1}}{2}}|\phi_\a\lb. %\no
%&=e^{i(L_0+\bar{L}_0)t}e^{\mathchar`-
% \rho(e^{\mathchar`-i\phi}\frac{L_{1}-\bar{L}_{\mathchar`-1}}{2}+e^{i\phi}\frac{\bar{L}_{1}-L_{\mathchar`-1}}{2})}|\phi_\a\lb.
\label{bulklocal}
\eal
This consists of states in a conformal family of $\CO_\a$, thus it is an eigenstate of the conformal Casimir $L^2=L^2_0-(L_1L_{-1}+L_{-1}L_{1})/2+(L\r \bar{L})$
with the eigenvalue $\Delta_\a(\Delta_\a-2)=m^2_\a R^2$. The Casimir operator acts on the bulk local states (\ref{bulklocal}) as the d'Alembertian operator in the global AdS coordinate, thus we can confirm that they satisfy free field equations of motions in the global AdS (\ref{eq}).
One can also see that two point functions of the bulk local states become bulk-to-bulk propagators in the global coordinate \cite{cMERA}
\ba
\la \phi_\a(\rho,\phi,t) \phi_\a(\rho',\phi',t')\lb=\f{1}{2\s{\sigma^2-1}(\sigma+\s{\sigma^2-1})^{\Delta_\a-1}}=\f{e^{-(\Delta_\a-1)D}}{2\sinh D},
\ea
where $\sigma=\cosh D$ is an AdS-invariant function which is given by
\ba
\sigma(\boldsymbol{x}|\boldsymbol{x}')=\cosh\rho\cosh\rho'\cos\Delta t-\sinh\rho\sinh\rho'\cos\Delta\phi.
\ea
The expression (\ref{bulklocal}) is determined by the symmetry and it seems that large $c$ plays no role in the construction. However, it is necessary for reproducing the Fock space like structure of the Hilbert space of AdS. In the large $c$ limit, local operators in CFT  behave as generalized free fields whose correlators factorize and this reproduces the property of correlators of free local fields in AdS spacetime.
%\ba
%\la \CO_1\ddd\CO_{n}\lb=\la \CO_1\CO_{2}\lb\ddd\la \CO_{n-1}\CO_{n}\lb+{\rm permutations\q up \ to\ }\CO(1/c).
%\e
\\
\\
\textbf{Map between the boundary and the bulk points}\\
In the rest of this subsection, we will see how the $SL(2,\mathbb{R})$ transformation $g(\rho,\phi,t)$ acts on the bulk local state $|\phi_\a\lb$ in more detail
\ba
g(\rho,\phi,t)=%e^{i(L_0+\bar{L}_0)t}e^{\mathchar`-
%\rho(e^{\mathchar`-i\phi}\frac{L_{1}-\bar{L}_{\mathchar`-1}}{2}+e^{i\phi}\frac{\bar{L}_{1}-L_{\mathchar`-1}}{2})}.
e^{i(L_0+\bar{L}_0)t}e^{i(L_0-\bar{L}_0)\phi}e^{\mathchar`-
\rho\frac{L_{1}-\bar{L}_{\mathchar`-1}+\bar{L}_{1}-L_{\mathchar`-1}}{2}}.
\label{slgl}
\ea
We first argue about the translation in the spatial directions of the bulk coordinates $(\rho,\phi)$; $g(\rho,\phi)=%e^{\mathchar`-
%\rho(e^{\mathchar`-i\phi}\frac{L_{1}-\bar{L}_{\mathchar`-1}}{2}+e^{i\phi}\frac{\bar{L}_{1}-L_{\mathchar`-1}}{2})}
e^{i(L_0-\bar{L}_0)\phi}e^{\mathchar`-
\rho(\frac{L_{1}-\bar{L}_{\mathchar`-1}+\bar{L}_{1}-L_{\mathchar`-1}}{2})}$ and its action on a primary operator $\CO_\a$ inserted at the origin of the Euclidean plane $(z,\z)$.
Under the conformal transformation $g(\rho,\phi)$, the origin will move to a point $z_0$ on the Euclidean boundary:
\ba
z_{0}=\tanh\f{\rho}{2}e^{i\phi},\ \z_{0}=\tanh\f{\rho}{2}e^{-i\phi}.\label{map0}
\ea
On the bulk side, $g(\rho,\phi)$ moves a bulk point from the origin of AdS to the point $(\rho,\phi)$. Thus the primary operator inserted at $(z_0,\z_0)$ creates an excitation around $(\rho,\phi,t=0)$. In this way, we can construct the map between the boundary points where primary operators are inserted and the bulk points around which the dual bulk excitations exist.
\begin{figure}
\begin{center}
  \includegraphics[width=4.5cm]{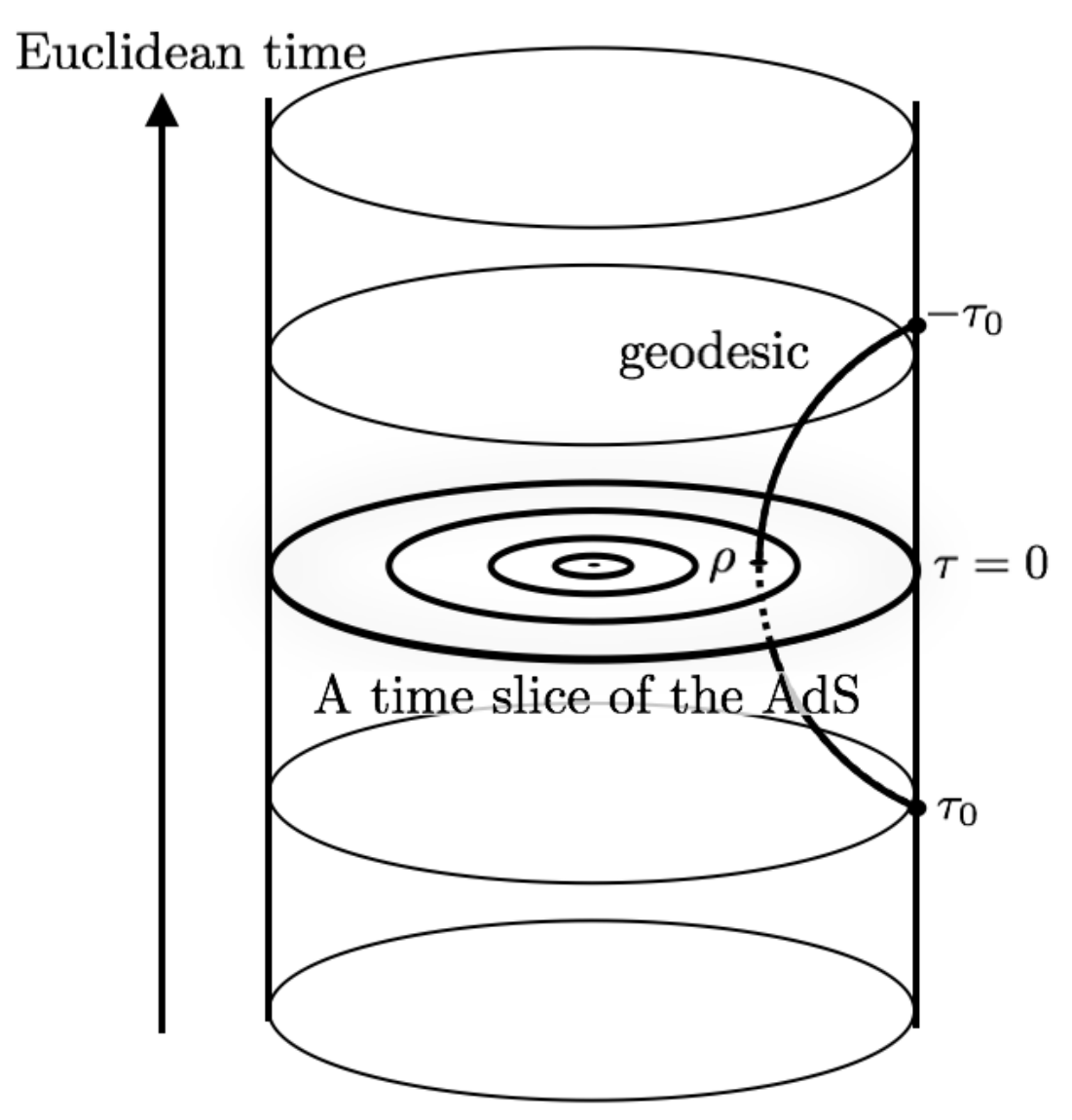}
  \caption{The bulk position $\rho$ can be obtained as an intersection of the timeslice $\tau=0$ and a geodesic which connects the boundary points $(\phi,\tau_0)$ and $(\phi,-\tau_0)$.}
  \label{geodesic3}
 \end{center}
\end{figure}
This map can be obtained from the geodesics as we can confirm in the viewpoint of $SL(2,\mathbb{R})$ symmetry;
the bulk position $\rho$ is obtained as an intersection of the timeslice $\tau=0$ and a geodesic which connects the boundary points $(\phi,\tau_0)$ and $(\phi,-\tau_0)$, see Figure \ref{geodesic3}. That is to say, the primary operator inserted at $(z_0,\z_0)$ creates the bulk excitation at $t=0$ centered around the geodesic which ends at $(z_0,\z_0)$\footnote{We expect that our prescription may have a connection with the correspondence between OPE blocks and bulk local operators smeared along the geodesic which is  proposed recently in \cite{CLMMS} while the explicit check of its connection remains to be done.}.
More generally,under the conformal transformation $g(\rho,\phi,t)$, the origin will move to a boundary point $z_0$;
\ba
z_{0}=\tanh\f{\rho}{2}e^{i(t+\phi)},\ \z_{0}=\tanh\f{\rho}{2}e^{i(t-\phi)}.\label{map}
\ea
 We would also like to mention that this map was also naturally obtained in the emergent AdS space from an optimization of Euclidean path-integral as discussed in \cite{OPT}. The original path-integral for $-\infty<\tau<0$ to produce the ground state of the CFT, which is a half of the boundary cylinder gets optimized into a round disk. This disk is the hyperbolic space which is identified the bulk time slice $\tau=0$. Under this correspondence a boundary point is mapped into a boundary point just as in Figure \ref{geodesic3}.

For descendant states, we need additional information about how the energy momentum tensor transforms under $g(\rho,\phi,t)$. The conformal transformation generates the following coordinate transformation
% \bal
%z'=\f{z+z_{0}}{1+e^{-2i(t+\phi)}z_{0}z}%-z_{0}=-2\f{\cosh\rho z-\sinh\rho}{\sinh\rho z-\cosh\rho-1}
% ,\
 %\z'=\f{\z+\z_{0}}{1+e^{-2i(t-\phi)}\z_{0}\z} .
% \eal
  \bal
z'=f(z)=\f{z+\tanh\f{\rho}{2}}{1+z\tanh\f{\rho}{2}}e^{i(t+\phi)}%-z_{0}=-2\f{\cosh\rho z-\sinh\rho}{\sinh\rho z-\cosh\rho-1}
 ,\
\z'=\tilde{f}(z)=\f{\z+\tanh\f{\rho}{2}}{1+\z\tanh\f{\rho}{2}}e^{i(t-\phi)}.
 \eal
Then under the global conformal transformation $g(\rho,\phi,t)$, the energy momentum tensor transforms as
 \ba
 g(\rho,\phi,t)T(z)g(\rho,\phi,t)^{-1}=f^{*}T(z)\equiv\biggl(\f{\pp z'}{\pp z}\biggl)^{2}T(f(z)) \no
 g(\rho,\phi,t)\bar{T}(\z)g(\rho,\phi,t)^{-1}=\tilde{f}^{*}\bar{T}(\z)\equiv\biggl(\f{\pp \z'}{\pp \z}\biggl)^{2}\bar{T}(\tilde{f}(z)).
 \ea
By using these expressions for the energy momentum tensor, we define new Virasoro ( or $SL(2,\mathbb{R})$) generators around $(z_0,\z_0)$ as follows
\bal
L^{z_0}_{n}\CO(z_{0})&=\oint_{z_{0}}\f{dz}{2\pi i}z^{n+1}f^{*}T(z)\CO(z_{0})%=\oint_{z_{0}}\f{dz}{2\pi i}z^{n+1}\biggl(\f{\pp z'}{\pp z}\biggl)^{2}T(f(z))\CO(z_{0})
\no
\bar{L}^{\z_0}_{n}\CO(\z_{0})&=\oint_{\z_{0}}\f{d\z}{2\pi i}\z^{n+1}\tilde{f}^{*}\bar{T}(\z)\CO(\z_{0}).%=\oint_{\z_{0}}\f{d\z}{2\pi i}\z^{n+1}\biggl(\f{\pp \z'}{\pp \z}\biggl)^{2}\bar{T}(\tilde{f}(\z))\CO(\z_{0})\no.
\eal

%We can calculate these generators as
%\bal
%L^{z_0}_{n}&=\oint_{0} dz'\biggl(\f{\pp z'}{\pp z}\biggl)(z(z'))^{n+1}T(z')\no&=
%\oint_{0}dz'\f{(1-e^{-2i(t+\phi)}z_0z')^{1-n}(z'-z_0)^{n+1}}{1-e^{-2i(t+\phi)}z_0^2}T(z')\no
%\LR L^{z_0}_{0}&=\cosh \rho L_0-\f{\sinh \rho}{2}(e^{-i(t+\phi)}L_1+e^{i(t+\phi)}L_{-1})=\no
%L^{z_0}_{\pm1}&=\cosh^2\f{\rho}{2}L_{\pm1}-e^{\pm i(t+\phi)}\sinh\rho L_{0}+e^{\pm 2i(t+\phi)}\sinh\f{\rho}{2}L_{-1}
%\eal
%Thus we can identify
%\bal
%L_{0}^{z_{0}}&=e^{-\f{\rho}{2}(e^{-i(t+\phi)}L_1-e^{i(t+\phi)}L_{-1})}L_0 e^{\f{\rho}{2}(e^{-i(t+\phi)}L_1-e^{i(t+\phi)}L_{-1})}\no &=\cosh \rho L_0-\f{\sinh \rho}{2}(e^{-i(t+\phi)}L_1+e^{i(t+\phi)}L_{-1})\no
%L_{\pm 1}^{z_{0}}&=e^{-\f{\rho}{2}(e^{-i(t+\phi)}L_1-e^{i(t+\phi)}L_{-1})}L_{\pm 1} e^{\f{\rho}{2}(e^{-i(t+\phi)}L_1-e^{i(t+\phi)}L_{-1})}\no &=e^{\pm i(t+\phi)}(-\sinh \rho L_0+\f{\cosh \rho}{2}(e^{-i(t+\phi)}L_1+e^{i(t+\phi)}L_{-1})\pm \f{1}{2}(e^{-i(t+\phi)}L_1-e^{i(t+\phi)}L_{-1})).
%\eal

From the argument we discussed above, we know how the primary states and descendant states transform under the conformal transformation $g(\rho,\phi,t)$. They are expressed as operators inserted at the boundary points $(z_0,\z_0)$ which correspond to the bulk points $(\rho,\phi,t)$ through the map (\ref{map}).
From the arguments above, we can see that the bulk local states is obtained by summing up a primary operator and its ``descendants'' inserted at the boundary point $(z_0,\z_0)$\footnote{Here we omit the normalization constant $(\f{\pp z'}{\pp z})^{h_{\a}}|_{z_0}(\f{\pp \z'}{\pp \z})^{\bar{h}_{\a}}|_{\z_0}$ which comes from the conformal transformation of the primary operator. In this paper, we implicitly include such factors in the expressions of the primaries $\CO_\a$.}:
\bal
|\phi_\a(\rho,\phi,t)\lb
&=\sum_{k=0}^{\infty}(-1)^k\f{\Gamma(\Delta)}{k!\Gamma(k+\Delta)}%(-1)^k
(L_{-1}^{z_{0}})^{k}(\bar{L}^{\z_{0}}_{-1})^{k} \CO_\a(z_0,\z_0)|0\lb\no
z_{0}&=\tanh\f{\rho}{2}e^{i(t+\phi)},\ \z_{0}=\tanh\f{\rho}{2}e^{i(t-\phi)}.
\eal
\textbf{Summary of the construction of bulk local states}\\
Here we summarize our method of constructing bulk local states. This method relies on the map between the bulk points and the boundary points. This map can be calculated by the geodesics which connect the bulk points and boundary points.  We insert the primary operator $\CO_\a$ dual to $\phi_\a$ at the boundary point $z_0$ which creates the bulk excitation centered around the bulk point corresponding to $z_0$ via the map:
\ba
\CO_\a(z_0)|0\lb.
\ea
We define the Virasoro generators $L^{z_0}_n,\bar{L}^{z_0}_n$ around  $(z_0,\bar{z}_0)$ as
\bal
L^{z_0}_{n}\CO(z_{0})&=\oint_{z_{0}}\f{dz}{2\pi i}z^{n+1}f^{*}T(z)\CO(z_{0})%=\oint_{z_{0}}\f{dz}{2\pi i}z^{n+1}\biggl(\f{\pp z'}{\pp z}\biggl)^{2}T(f(z))\CO(z_{0})
\no
\bar{L}^{\z_0}_{n}\CO(\z_{0})&=\oint_{\z_{0}}\f{d\z}{2\pi i}\z^{n+1}\tilde{f}^{*}\bar{T}(\z)\CO(\z_{0}),%=\oint_{\z_{0}}\f{d\z}{2\pi i}\z^{n+1}\biggl(\f{\pp \z'}{\pp \z}\biggl)^{2}\bar{T}(\tilde{f}(\z))\CO(\z_{0})\no.
\eal
where $f(z),\tilde{f}(\z)$ are conformal transformations corresponding to $g(\rho,\phi,t)$
 \bal
z'=f(z)=\f{z+\tanh\f{\rho}{2}}{1+z\tanh\f{\rho}{2}}e^{i(t+\phi)}%-z_{0}=-2\f{\cosh\rho z-\sinh\rho}{\sinh\rho z-\cosh\rho-1}
 ,\
\z'=\tilde{f}(\z)=\f{\z+\tanh\f{\rho}{2}}{1+\z\tanh\f{\rho}{2}}e^{i(t-\phi)}.
 \eal
In order to localize the bulk excitation and obtain the expression for the bulk local state, we dress the primary operator with the ``localizing operator'' $\hat{K}^{\boldsymbol{x}}=\sum_{k}(-1)^k\f{\Gamma(\Delta)}{k!\Gamma(k+\Delta)}%(-1)^k
(L_{-1}^{\boldsymbol{x}})^{k}(\bar{L}^{\boldsymbol{x}}_{-1})^{k}$
\bal
|\phi_\a(\rho,\phi,t)\lb
&=\sum_{k=0}^{\infty}(-1)^k\f{\Gamma(\Delta)}{k!\Gamma(k+\Delta)}%(-1)^k
(L_{-1}^{z_{0}})^{k}(\bar{L}^{\z_{0}}_{-1})^{k} \CO_\a(z_0,\z_0)|0\lb.
\eal
We can reconstruct bulk local states with this method even in the Rindler coordinate and BTZ black hole spacetimes  as we will see below.% In appendix A, we do so in the Poincare coordinate. 
\subsection{Rindler-AdS}\label{Rindler}
In this section, we consider the Rindler coordinate which is useful for understanding BTZ black hole spacetime. Rindler coordinate consists of four wedges: right/left/future and past wedge. The relations with the global coordinate are:
\begin{alignat}{4}
\cosh\rho \cos t&=\cosh\rho_r\cosh\phi_r&=&\q \cosh\rho_l\cosh\phi_{l}&=&\cos\rho_f\cosh\phi_f&=&\q \cos\rho_p\cosh\phi_p\no
\cosh\rho \sin t&=\sinh\rho_r\sinh t_r&=&\q\sinh\rho_l\sinh t_l&=&\sin\rho_f\cosh t_f&=&-\sin\rho_p\cosh t_p\no
\sinh\rho \sin \phi&=\cosh\rho_r\sinh \phi_r&=&-\cosh\rho_l\sinh \phi_{l}&=&\cos\rho_f\sinh \phi_f&=&\q\cos\rho_p\sinh \phi_p\no
\sinh\rho \cos \phi&=\sinh\rho_r\cosh t_r&=&-\sinh\rho_l\cosh t_{l}
&=&\sin\rho_f\sinh t_f&=&\q\sin\rho_p\sinh t_p,
\end{alignat}
where we write coordinates of the right, left, future and past wedge as $(\rho_r,\phi_r,t_r), (\rho_l,\phi_l,t_l)$, $(\rho_f,\phi_f,t_f)$ and $(\rho_p,\phi_p,t_p)$ respectively and $r=R\cosh\rho_r=R\cosh\rho_l=R\cos\rho_f=R\cos\rho_p$.
The metric of the right wedge is written as
\bal
ds^{2}&=-(r^{2}-R^{2})dt_{r}^{2}+\f{R^{2}}{r^{2}-R^{2}}dr^{2}+r^{2}d\phi_{r}^{2}\no
&=R^{2}(-\sinh^{2}\rho_{r} dt_{r}^{2}+d\rho_{r}^{2}+\cosh^{2}\rho_{r}
d\phi_r^{2})\no
&r>R\ (\rho_r>0),-\infty<t_r<\infty,-\infty<\phi_r<\infty.\eal
The left wedge is related to the right wedge by the analytic continuation $t_{l}=-t_{r}+i\pi$, $\phi_{l}=-\phi_{r}$\footnote{Here we define the left coordinate $(\phi_l,t_l)$ with $-1$ flip to take the total Hamiltonian as $H_{\rm total}=H_l+H_r$.}. %or $t_{l}=-t_{r}$,$\rho_{l}=-\rho_{r}$, $\phi_{l}=-\phi_{r}$
The metric of the future wedge is written as
\bal
ds^{2}&=(R^{2}-r^{2})dt_f^{2}-\f{R^{2}}{R^{2}-r^{2}}dr^{2}+r^{2}d\phi_f^{2}\no
&=R^{2}(\sin^{2}\rho_f dt_f^{2}-d\rho_f^{2}+\cos^{2}\rho_f
d\phi_f^{2})\no
&r<R\ (\rho_f>0),-\infty<t_f<\infty,-\infty<\phi_f<\infty\eal
One can enter from the right wedge to the future wedge by the analytic continuation $t_f=t_r+i\pi/2$.\\
The $SL(2,\mathbb{R})$ generators can be expressed as differential operators in the right/left wedge:
\bal
L_0^{r,l}&=-\pp_+\no
L_{\pm 1}^{r,l}&=-\f{\sqrt{r^2-R^2}}{2r}e^{\pm x^{+}}\biggl[ \f{2r^2-R^2}{r^2-R^2}\pp_+-\f{R^2}{r^2-R^2}\pp_-\mp r\pp_r \biggl],\no
\bar{L}_0^{r,l}&=-\pp_-\no
\bar{L}_{\pm 1}^{r,l}&=-\f{\sqrt{r^2-R^2}}{2r}e^{\pm x^{-}}\biggl[\f{2r^2-R^2}{r^2-R^2}\pp_--\f{R^2}{r^2-R^2}\pp_+\mp r\pp_r \biggl].\label{LR}
\eal
while $SL(2,\mathbb{R})$ symmetry is not globally defined in each wedge due to the existence of the horizon. In (\ref{LR}), we defined a light-cone coordinate in the Rindler spacetime
\ba
x^{\pm}\equiv\phi_{r,l}\pm t_{r,l}.
\ea
Notice that we defined a light-cone coordinate in the global-AdS as $x^{\pm}\equiv t\pm \phi$.
%These are expressed in the hyperbolic basis (see Appendix A) while in the global coordinate we considered in the elliptic basis.We will see the relation between these bases later.
\begin{figure}
\begin{center}
  \includegraphics[width=12cm]{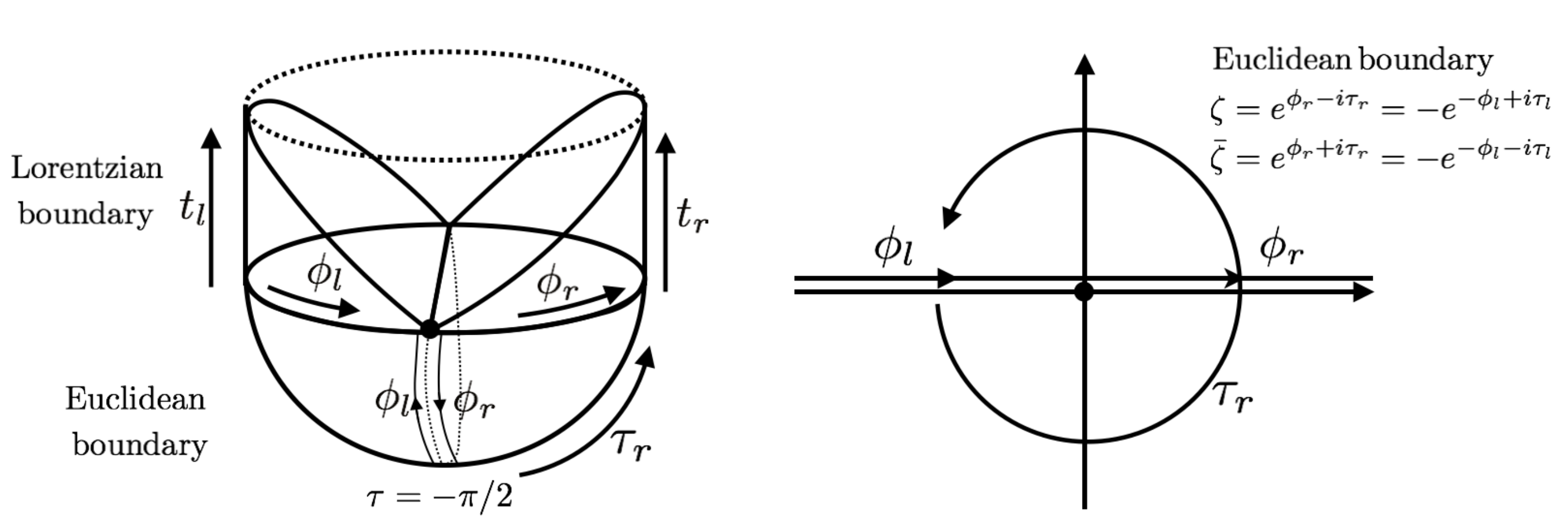}
  \caption{Rindler coordinate and Euclidean plane defined on its boundary}
 \end{center}
\end{figure}
We will see below that the combinations of $L^{r,l}_n$ make generators of globally defined $SL(2,\mathbb{R})$ symmetry which act on both left and right wedge.

We introduce an Euclidean coordinate on the boundary of the Rindler-AdS:
\ba
\zeta=e^{x^{+}}=e^{\phi_r-i \tau_r}=-e^{-\phi_l+i \tau_l},\q \bar{\zeta}=e^{x^{-}}=e^{\phi_r+ i \tau_r}=-e^{-\phi_l-i \tau_l}.
\ea
where $\tau_{r,l}$ is the Euclidean time defined as $\tau_{r,l}=it_{r,l}$.
The generators $L_{n}^{l}$ and $L_{n}^{r}$ are if combined, become globally defined conformal generators in the CFT:
\bal
L^{R}_{n}& \equiv
\int_{\mathbb{R}} d\zeta \zeta^{n+1}T(\zeta)=(\int_{\mathbb{R}_+} d\zeta+\int_{\mathbb{R}_-} d\zeta) \zeta^{n+1}T(\zeta)\no &=\int_{-\infty}^{\infty} d\phi_{r} e^{n \phi_{r}}T(\phi_{r})-(-1)^{n}\int_{-\infty}^{\infty} d\phi_{l} e^{-n \phi_{l}}T(\phi_{l})\no
&=L^{r}_{n}-(-1)^{n}L^{l}_{-n}\no
\bar{L}^{R}_{n}&\equiv
\int_{\mathbb{R}} d\bar{\zeta}\bar{\zeta}^{n+1}\bar{T}(\bar{\zeta})=(\int_{\mathbb{R}_+} d\bar{\zeta}+\int_{\mathbb{R}_-} d\bar{\zeta}) \bar{\zeta}^{n+1}\bar{T}(\bar{\zeta})\no &=\int_{-\infty}^{\infty} d\phi_{r} e^{n \phi_{r}}\bar{T}(\phi_{r})-(-1)^{n}\int_{-\infty}^{\infty} d\phi_{l} e^{n \phi_{l}}\bar{T}(\phi_{l})\no
&=\bar{L}^{r}_{n}-(-1)^{n}\bar{L}^{l}_{-n}.
\eal
That is to say, though the existence of the horizon destroys the $SL(2,\mathbb{R})$ symmetry of each wedge in Rindler spacetime,  it will be recovered if one take linear combinations of the generators which act on both wedges. Through the coordinate transformation between the global and Rindler:
\bal
\tanh(\f{\phi_r\pm t_r}{2})&=\tan(\f{\phi\pm t}{2})\ \no\LR\  \zeta=i\f{1-iz}{1+iz},\ \bar{\zeta}=-i\f{1+i\z}{1-i\z},&\q z=-i\f{1+i\zeta}{1-i\zeta},\ \bar{z}=i\f{1-i\bar{\zeta}}{1+i\bar{\zeta}},\label{glri}
\eal
we can find the relations between $L^{R}_n$ and $L_n$ defined in the global coordinate
%\begin{alignat}{4}
%L_0&=-i\f{L^{\rm Rin}_1+L^{\rm Rin}_{-1}}{2}
%&L_{\pm 1}&=-iL^{\rm Rin}_0\pm \f{L^{\rm Rin}_1-L^{\rm Rin}_{-1}}{2}
%&\bar{L}_0&=i\f{\bar{L}^{\rm Rin}_1+\bar{L}^{\rm Rin}_{-1}}{2}
%5&\bar{L}_{\pm 1}&=i\bar{L}^{\rm Rin}_0\pm\f{\bar{L}^{\rm Rin}_1-\bar{L}%^{\rm Rin}_{-1}}{2}, \no
%L^{\rm Rin}_0&=i\f{L_1+L_{-1}}{2}
%&L^{\rm Rin}_{\pm 1}&=iL_0\pm\f{L_1-L_{-1}}{2}
%&\bar{L}^{\rm Rin}_0&=-i\f{\bar{L}_1+\bar{L}_{-1}}{2}
%&\bar{L}^{\rm Rin}_{\pm 1}&=-i\bar{L}_0\pm\f{\bar{L}_1-\bar{L}_{-1}}{2}.
%\end{alignat}
\begin{alignat}{2}
L_0&=-i\f{L^{R}_1+L^{R}_{-1}}{2},\
&L_{\pm 1}&=-iL^{R}_0\pm \f{L^{R}_1-L^{R}_{-1}}{2}, \no
\bar{L}_0&=i\f{\bar{L}^{R}_1+\bar{L}^{R}_{-1}}{2},\
&\bar{L}_{\pm 1}&=i\bar{L}^{R}_0\pm\f{\bar{L}^{R}_1-\bar{L}^{R}_{-1}}{2}, \no
L^{R}_0&=i\f{L_1+L_{-1}}{2},\
&L^{R}_{\pm 1}&=iL_0\pm\f{L_1-L_{-1}}{2}, \no
\bar{L}^{R}_0&=-i\f{\bar{L}_1+\bar{L}_{-1}}{2},\
&\bar{L}^{R}_{\pm 1}&=-i\bar{L}_0\pm\f{\bar{L}_1-\bar{L}_{-1}}{2}.
\end{alignat}\\
\textbf{Thermofield double description}\\
In the Rindler coordinate, states in CFT are defined by the Euclidean path-integrals over $ -\infty<\phi_r<\infty,-\pi<\tau_r<0$. The Hilbert space of the CFT defined on Im$(\zeta)=0$ is naturally decomposed as the direct product of the Hilbert space of the left CFT:${\cal H}_l$ and the right CFT: ${\cal H}_r$
\ba
{\cal H}_{\rm CFT}\simeq{\cal H}_l\otimes {\cal H}_r.
\ea
The ``vacuum'' state defined by the Euclidean path integral without any operator insertions is expressed as a thermofield double state.%\footnote{%Since we defined time in the left CFT with (-1) flip relative to   the right CFT, thus
%We need a time-reversal operator ${\cal T}$ which acts on the left energy eigenstates:$|\Psi_{\rm TFD}\lb=\sum_{E}e^{-\beta E/2}{\cal T}|E\lb_{l}\otimes |E\lb_{r}$.In this paper we write ${\cal T}|E\lb_l$ just as $|E\lb_l$ for simplicity.%If the CFT is invariant under time-reversal and you choose ${\cal T}$ as ${\cal T}|E\lb_l=|E\lb_l$,we get the expression (\ref{tfd}).
%}
\ba
|\Psi_{\rm TFD}\lb=\sum_{E}e^{-\beta E/2}|E\lb_{l}\otimes|E\lb_{r},\label{tfd}
\ea
where $|E\lb_{l,r}$ represent energy eigenstates of each CFT and $\beta=2\pi$ is the period of the Euclidean Rindler time. Thermofield double state is dual to the Rindler spacetime without any excitations. Notice that it has a $SL(2,\mathbb{R})$ symmetry:
\ba
L^{R}_n|\Psi_{\rm TFD}\lb=\bar{L}^{R}_n|\Psi_{\rm TFD}\lb=0\LR (L^{r}_{n}-(-1)^{n}L^{l}_{-n})|\Psi_{\rm TFD}\lb=(\bar{L}^{r}_{n}-(-1)^{n}\bar{L}^{l}_{-n})|\Psi_{\rm TFD}\lb=0\nonumber,
\ea
for $n=-1,0,1$.\\ \\
\textbf{Construction of the bulk local states from the TFD states}\\
We move on to the construction of a bulk local state in the black hole dual to the thermofield double state. It is defined as $|\phi_\a(\boldsymbol{x})\lb_{\rm Rindler}\equiv \hat{\phi}^{\rm CFT}_\a(\boldsymbol{x})|\Psi_{\rm TFD}\lb$ where $\hat{\phi}^{\rm CFT}_\a(\boldsymbol{x})$ is a CFT operator dual to a free scalar field $\phi_{\a}(\boldsymbol{x})$ in the Rindler-AdS coordinate. We mainly focus on the reconstruction of the right wedge while the same argument is possible in the left wedge of the Rindler coordinate. The strategy is the same as the global coordinate as we will explain below.

\begin{figure}
\begin{center}
  \includegraphics[width=4.5cm]{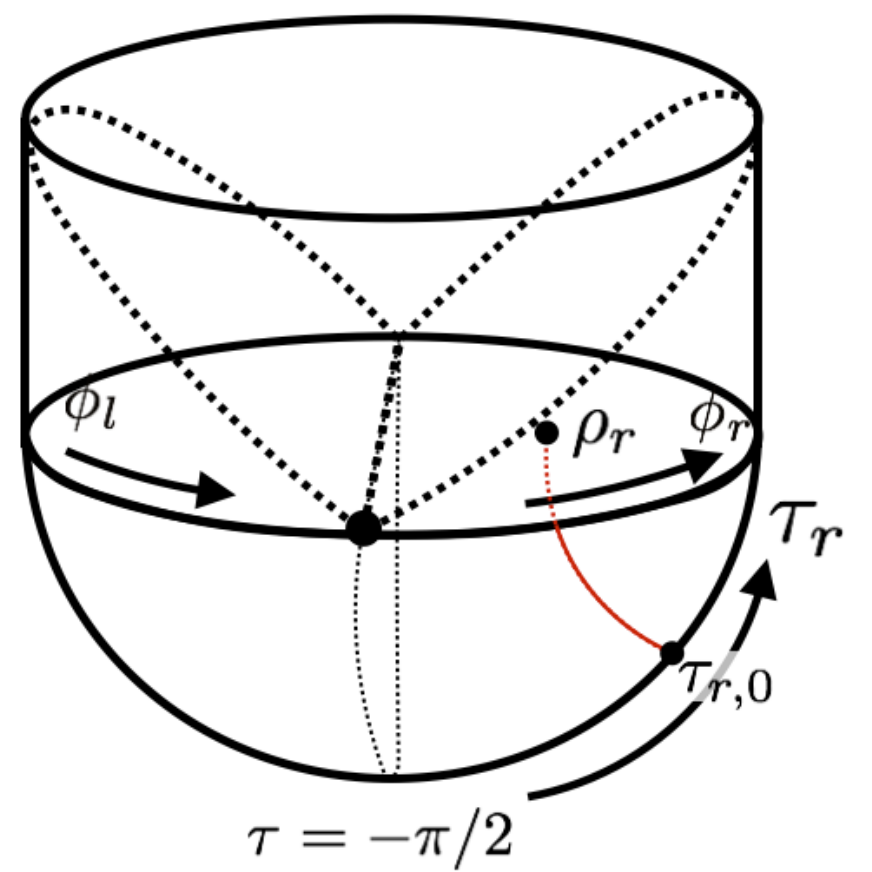}
  \caption{We can construct the map between bulk points and boundary points. This map can be obtained by the geodesics just the same as the global coordinate.}\label{fbtzma}
 \end{center}
\end{figure}

First let us consider the map between the bulk points and the boundary point. From the relation between the global and the Rindler coordinate (\ref{glri}), we can see that the boundary point $(z,\bar{z})=0$ corresponds to $(\zeta,\bar{\zeta})=(i,-i)\LR (\phi_r,\tau_r)=(0,-\pi/2)$ on the Euclidean boundary of the Rindler coordinate (see Figure \ref{fbtzma}). Since the origin of the Euclidean plane $(z,\bar{z})$ corresponds to the center of the AdS, the boundary point $(\phi_r,\tau_r)=(0,-\pi/2)$ corresponds to $(\rho_r,\phi_r,t_r)=(0,0,0)$ in the Rindler coordinate. Thus the excitation around the center of the AdS spacetime can be written as a state created by Euclidean path integral over $-\pi<\tau<0$ with a primary operator inserted at $(\phi_r,\tau_r)=(0,-\pi/2)$
\ba
\CO(0)|0\lb\simeq e^{-\beta H_r/4}\CO_r(0,0)e^{-\beta H_l/4}\sum_{E}|E\lb_l\otimes |E\lb_r=\CO_r(0,-\pi/2)|\Psi_{\rm TFD}\lb,\nonumber
\ea
where we write a primary operator in the right CFT as $\CO_{r}$.
Thus we can construct a bulk local state at the center of AdS as \footnote{In this subsection we omit the labels of primaries:$\a$ to simplify the notations.}
\bal
|\phi(0)\lb_{\rm Rindler}&=\sum_{k=0}^{\infty}(-1)^k\f{\Gamma(\Delta)}{k!\Gamma(k+\Delta)}%(-1)^k
L_{-1}^{k}\bar{L}_{-1}^{k}\CO_{r}(0,-\pi/2)|\Psi_{\rm TFD}\lb.
\eal
By construction, it satisfies the condition for the local scalar operator in AdS (\ref{condi})
\bal
(L_0-\bar{L}_0)|\phi(0)\lb_{\rm Rindler}&=(L_{\pm 1}+\bar{L}_{\mp 1})|\phi(0)\lb_{\rm Rindler}=0\no
\LR (L^{R}_0-\bar{L}^{R}_0)|\phi(0)\lb_{\rm Rindler}&=(L^{R}_{\pm 1}+\bar{L}^{R}_{\mp 1})|\phi(0)\lb_{\rm Rindler}=0.\label{Ishibashitfd}
\eal
\\
\textbf{$SL(2,\mathbb{R})$ transformation to different bulk points}\\
We can write the $SL(2,\mathbb{R})$ generators which move the bulk point from the center of the AdS to other points $(\rho_r,\phi_r,t_r)$ as
\bal
g(\rho_r,\phi_r,t_r)&=e^{(L^{R}_0-\bar{L}^{R}_0)t_r}e^{(L^{R}_0+\bar{L}^{R}_0)\phi_r}e^{-
\rho_r\f{L^{R}_1-L^{R}_{-1}+\bar{L}^{R}_1-\bar{L}^{R}_{-1}}{2}}\no&=e^{i\f{L_{1}+L_{-1}+\bar{L}_{1}+\bar{L}_{-1}}{2}t_{r}}e^{i\f{L_{1}+L_{-1}-\bar{L}_{1}-\bar{L}_{-1}}{2}\phi_{r}}
e^{-\rho_{r}\f{L_{1}-L_{-1}+\bar{L}_{1}-\bar{L}_{-1}}{2}}.\label{eq:rinslq}
\eal
We first consider the bulk point $(\rho_r,0,0)$. It corresponds to the boundary point $z_0$:
\bal
z_0=e^{\tau_0+i\phi_0}=\tan(\f{\tau_{r,0}+\pi/2}{2})=\tanh\f{\rho_r}{2}.
\eal
Thus we can see that the bulk excitation around $(\rho_
r,\phi_r,t_r)$ is dual to the CFT state
\ba
\CO(z_0,\z_0)|0\lb\simeq \CO_r(\phi_r,\gamma)|\Psi_{\rm TFD}\lb,\nonumber
\ea
where we introduce a parameter $\gamma=t_r-i\tau_{r,0}$ which parametrizes a complex time in the boundary coordinate of the Rindler-AdS and $z_0,\z_0$ are defined as
\ba
z_0=\f{\tanh\f{\rho_r}{2}+i\tanh\f{\phi_r+t_r}{2}}{1-i\tanh\f{\rho_r}{2}\tanh\f{\phi_r+t_r}{2}},\ \z_0=\f{\tanh\f{\rho_r}{2}-i\tanh\f{\phi_r-t_r}{2}}{1+i\tanh\f{\rho_r}{2}\tanh\f{\phi_r-t_r}{2}}.\label{Rightmap}
\ea

In general, the conformal transformations generated by (\ref{eq:rinslq}) 
are expressed as
\bal
z'&=f_{\boldsymbol{x}}(z)=\f{(1+i\tanh\f{\rho_r}{2}\tanh\f{\phi_r+t_r}{2})z+\tanh\f{\rho_r}{2}+i\tanh\f{\phi_r+t_r}{2}}{(\tanh\f{\rho_r}{2}-i\tanh\f{\phi_r+t_r}{2})z+1-i\tanh\f{\rho_r}{2}\tanh\f{\phi_r+t_r}{2}},\no
\z'&=\tilde{f}_{\boldsymbol{x}}(\z)=\f{(1-i\tanh\f{\rho_r}{2}\tanh\f{\phi_r-t_r}{2})\z+\tanh\f{\rho_r}{2}-i\tanh\f{\phi_r-t_r}{2}}{(\tanh\f{\rho_r}{2}+i\tanh\f{\phi_r-t_r}{2})\z+1+i\tanh\f{\rho_r}{2}\tanh\f{\phi_r-t_r}{2}}.\label{Right}
\eal
We define the Virasoro generators transformed by the conformal transformation $f_{\boldsymbol{x}}$, $\tilde{f}_{\boldsymbol{x}}(\z)$ around the boundary point $(z_0,\z_0)\LR (\phi_r,\gamma)$ as
\bal
L^{{\boldsymbol{x}}}_n\CO(z_0)&=\oint_{z_0} dz z^{n+1}f_{\boldsymbol{x}}^{*}T(z)\CO(z_0),\no
\bar{L}^{{\boldsymbol{x}}}_n\CO_r(z_0)&=\oint_{\z_0} d\z \z^{n+1}\tilde{f}_{\boldsymbol{x}}^*\bar{T}(\z)\CO(\z_0).\label{RVira}
\eal
The map (\ref{Rightmap}) and the Virasoro generators defined around $(z_0,\z_0)$ (\ref{RVira}) enable us to write the bulk local states at $(\rho_r,\phi_r,t_r)$ in the right Rindler coordinate:
\bal
|\phi(\rho_r,\phi_r,t_r)\lb_{\rm Rindler}&=\sum_{k=0}^{\infty}(-1)^k\f{\Gamma(\Delta)}{k!\Gamma(k+\Delta)}(L^{{\boldsymbol{x}}}_{-1})^{k}(\bar{L}^{{\boldsymbol{x}}}_{-1})^{k}\CO_r(\phi_r,\gamma)|\Psi_{\rm TFD}\lb.\label{rightexpre}
\eal
The construction in the left wedge can be done in the same way as the right wedge.\\ \\
\textbf{Inside the horizon}\\
Next we consider the future wedge which corresponds to the inside the horizon of the Rindler coordinate.
The generators of the $SL(2,\mathbb{R})$ isometry can be obtained through the analytic continuation of the generators in the right (or left) wedge. We can find the $SL(2,\mathbb{R})$ transformation which moves a point in the future wedge can be expressed as
\bal
g(\rho_f,\phi_f,t_f)&=e^{(L^{R}_0-\bar{L}^{R}_0)t_f}e^{(L^{R}_0+\bar{L}^{R}_0)\phi_f}e^{
\rho_f\f{L^{R}_1+L^{R}_{-1}-\bar{L}^{R}_1-\bar{L}^{R}_{-1}}{2}}\no&=e^{i\f{L_{1}+L_{-1}+\bar{L}_{1}+\bar{L}_{-1}}{2}t_{f}}
e^{i\f{L_{1}+L_{-1}-\bar{L}_{1}-\bar{L}_{-1}}{2}\phi_{f}}e^{i(L_{0}+\bar{L}_0)\rho_{f}}.\label{eq:rinsl}
\eal
%From the expression above, we can see that the generator of the $\rho_f$ direction is equivalent to the Hamiltonian $H_G$ in the global coordinate.We place a primary operator $\CO_r$ at $(\phi_r,\tau_r)=(0,-\pi/2+\delta)$ which is dual to the excitation around the center of AdS. The bulk point $(\rho_f,0,0)$ corresponds to the boundary point $(t,\phi)=(\rho_f,0)$ where $t$ is the Lorentzian time of the global coordinate.By the map (\ref{glri}),the bulk direction $\rho_f$ and the Lorentzian time $t_{r,0}$ on the boundary in the Rindler coordinate are related as
%\bal
%r=\cos\rho_f=\f{1}{\cosh t_{r,0}}.
%\eal
%We can see that the limit $r\r0$ corresponds to $t_r\r \pm\infty$ on the boundary.
From the expression above, we can see that the generator of the $\rho_f$ direction is equivalent to the Hamiltonian $H_G$ in the global coordinate. A primary operator inserted at the origin of the $(z,\z)$ coordinate is invariant under the time evolution by $H_G$. Similarly the generator of the $\rho_f$ direction does not move the point where the primary operator inserted;$(\phi_r,\tau_r)=(0,-\pi/2)$. More generally, we can see that the bulk excitation around $(\rho_f,\phi_f,t_f)$ is dual to the CFT state
\ba
\CO(z_0)|0\lb\simeq \CO_r(\phi_f,\gamma)|\Psi_{\rm TFD}\lb.
\ea
where $\gamma=t_f+
i\pi/2$.
Though boundary points where the primary operators are inserted are independent of the bulk coordinate $\rho_f$, Virasoro generators defined through the conformal transformation (\ref{eq:rinsl}) depend non-trivially on $\rho_f$.
The conformal transformations generated by (\ref{eq:rinsl}) are expressed as
\bal
z'&=f_{\boldsymbol{x}}(z)=\f{z_{0}+ze^{i\rho_f}}{1-z_{0}ze^{i\rho_f}},\ z_{0}=i\tanh(\f{\phi_f+t_f}{2})\no
\z'&=\tilde{f}_{\boldsymbol{x}}(\z)=\f{\z_{0}+\z e^{i\rho_f}}{1-\z_{0}\z e^{i\rho_f}},\ \z_{0}=-i\tanh(\f{\phi_f-t_f}{2})
.\label{Future}
\eal
We define Virasoro generators conformally transformed by $f_{\boldsymbol{x}}, \tilde{f}_{\boldsymbol{x}}$ around $(z_0,\z_0)\LR (\phi_f,\gamma)$
\bal
L^{{\boldsymbol{x}}}_n\CO(z_0)&=\oint_{z_0} dz z^{n+1}f_{\boldsymbol{x}}^{*}T(z)\CO(z_0),\no
\bar{L}^{{\boldsymbol{x}}}_n\CO(z_0)&=\oint_{\z_0} d\z \z^{n+1}\tilde{f}_{\boldsymbol{x}}^*\bar{T}(\z)\CO(\z_0)\label{Vira}.
\eal
From the map (\ref{Future}) and the Virasoro generators defined in (\ref{Vira}), we can construct bulk local states at $(\rho_f,\phi_f,t_f)$ in the future Rindler wedge as follows:
\bal
|\phi(\rho_f,\phi_f,t_f)\lb_{\rm Rindler}&=\sum_{k=0}^{\infty}(-1)^k\f{\Gamma(\Delta)}{k!\Gamma(k+\Delta)} (L^{{\boldsymbol{x}}}_{-1})^{k}(\bar{L}^{{\boldsymbol{x}}}_{-1})^{k}\CO_r(\phi_f,\gamma)|\Psi_{\rm TFD}\lb.\label{futureexpre}
\eal
where $\gamma=t_f+i\pi/2$.

We constructed bulk local states both outside and inside the horizon in the Rindler coordinate.
By construction, the bulk local states satisfy the equations of motion in the Rindler coordinate
\ba
(\Box_{\rm Rindler}+m^2)|\phi(\boldsymbol{x})\lb_{\rm Rindler}=0.
\ea
We will see that the bulk local states we defined above exactly give the two point functions in the Rindler coordinate.
\\
\\
\textbf{Correlators}\\
Now let us consider correlators of the bulk local states such as
\ba
\la \Psi_{\rm TFD}|\phi(\rho,\phi,t)\phi(\rho',\phi',t')|\Psi_{\rm TFD}\lb.\label{correl}
\ea
First we consider the correlation between the bulk local operators in the right wedge. We can see that the expressions of the $SL(2,\mathbb{R})$ generators in the right Rindler wedge (\ref{eq:rinslq}) are the same as those of the global AdS coordinate (\ref{slgl}) with a replacement of the generators; $L_{n} \r L_{n}^R,\ \bar{L}_{n}\r \bar{L}_{n}^R$  and of the coordinates; $\phi\r -it_r,\ t\r -i\phi_r$.
Moreover, the bulk local states $\phi(\rho',\phi',t')|\Psi_{\rm TFD}\lb$ satisfy the ``localizing'' condition with respect to $(L_{n}^R,\bar{L}_{n}^R)$ (\ref{Ishibashitfd}). Thus calculations of the correlators (\ref{correl}) can be reduced to calculations of the correlators for the twisted Ishibashi states with insertions of the $SL(2,\mathbb{R})$ generators in the global coordinate
\ba
g(\rho,\phi,t)=e^{i(L_0+\bar{L}_0)t}e^{i(L_0-\bar{L}_0)\phi}e^{-
\rho\f{L_1-L_{-1}+\bar{L}_1-\bar{L}_{-1}}{2}},
\ea
with a 
replacement of parameters $\phi\r -it_r,\ t\r -i\phi_r$. That is to say, we can calculate the correlators in the Rindler coordinate as
\ba
\la \Psi_{\rm TFD}|\phi(\rho,\phi,t)\phi(\rho',\phi',t')|\Psi_{\rm TFD}\lb=\la \phi|g^{-1}(\rho_r,-it_r,-i\phi_r)g(\rho'_r,-it'_r,-i\phi'_r)|\phi\lb,\label{eqvt}
\ea
where $g(\rho_r,-it_r,-i\phi_r)$ means $g(\rho=\rho_r,\phi=-it_r,t=-i\phi_r)$ and $|\phi\lb$ is a twisted Ishibashi state (\ref{Ishibashi}). From the arguments above, we can check that two point functions of the bulk local states constructed from the thermofield double state reproduce the two point functions in the Rindler coordinate \cite{IS}
\ba
\la \Psi_{\rm TFD}|\phi(\rho,\phi,t)\phi(\rho',\phi',t')|\Psi_{\rm TFD}\lb=G^{\rm Rindler}(\rho,\phi,t;\rho',\phi',t'),\label{correl3}
\ea
where
\ba
G^{\rm Rindler}(\rho,\phi,t;\rho',\phi',t')=\f{1}{2\s{\sigma^2-1}(\sigma+\s{\sigma^2-1})^{\Delta-1}}.\label{correl4}
\ea
$\sigma$ is an AdS-invariant function given by
\bal
\sigma(\boldsymbol{x}_r|\boldsymbol{x}_r')&=\cosh\rho_r\cosh\rho_r'\cosh\Delta \phi-\sinh\rho_r\sinh\rho_r'\cosh\Delta t \no
&=\f{1}{R^2}[rr'\cosh\Delta \phi-\s{r^2-R^2}\s{r'^2-R^2}\cosh\Delta t],
\eal
where $\Delta \phi=\phi-\phi',\Delta t=t-t'$.
By the similar arguments, correlation functions between the bulk points in other wedges of the Rindler coordinate are written as (\ref{correl3})(\ref{correl4}) with AdS-invariant functions which is given by%\footnote{AdS-invariant functions in other wedges can also be written in the same manner.}
\bal
\sigma(\boldsymbol{x}_r|\boldsymbol{x}_f')&=\cosh\rho_f\cos\rho_f'\cosh\Delta \phi-\sinh\rho_f\sin\rho_f'\sinh\Delta t\no
&=\f{1}{R^2}[rr'\cosh\Delta \phi-\s{r^2-R^2}\s{R^2-r'^2}\sinh\Delta t]\no
\sigma(\boldsymbol{x}_f|\boldsymbol{x}_f')&=\cos\rho_f\cos\rho_f'\cosh\Delta \phi+\sin\rho_f\sin\rho_f'\cosh\Delta t\no
&=\f{1}{R^2}[rr'\cosh\Delta \phi+\s{R^2-r^2}\s{R^2-r'^2}\cosh\Delta t]\no
\sigma(\boldsymbol{x}_l|\boldsymbol{x}_r')&=\cosh\rho_l\cosh\rho_r'\cosh\Delta \phi+\sinh\rho_l\sinh\rho_r'\cosh\Delta t\no
&=\f{1}{R^2}[rr'\cosh\Delta \phi_r+\s{r^2-R^2}\s{r'^2-R^2}\sinh\Delta t],\label{sigma}
\eal
as examples.\\
We can also understand the equivalence between correlation functions for the bulk local operators on the thermofield double states and those on the vacuum states (\ref{eqvt}) from the Euclidean path-integral picture.
The bulk local states themselves are defined through the Euclidean path-integrals with operator insertions along different time directions, $\tau_r$ and $\tau$. However once we consider their correlators, they are both expressed as the Euclidean-path integrals over the two-dimensional sphere $\mathbb{S}^2$ with insertions of operators. Thus we find that the equivalence (\ref{eqvt}) holds for the correlators from the perspective of the Euclidean path-integral.

Notice that since higher point functions on the thermofield state in the large $c$ CFTs will factorize into two point functions in the leading order of the $1/c$ expansion, our construction is sufficient for the calculations of arbitrary higher point functions up to $1/c$ corrections.\\

\section{Bulk local states in the BTZ black holes}\label{BTZ}
In this section, we consider the bulk local state in the BTZ black holes.$SL(2,\mathbb{R})$ is no longer a globally defined symmetry in the black hole spacetimes. Our construction is based on the symmetry, thus one might think it is difficult to construct bulk local states in the black holes. However, it is well known that the  three dimensional black holes are locally equivalent to the pure AdS$_3$ spacetime and we can embed the black hole spacetimes into the Rindler coordinate of the pure AdS spacetime. Though the global structures of the black hole spacetimes and the pure AdS are different, we can define the bulk local states in the Rindler coordinate locally around a boundary point $(z_0,\z_0)$ as we saw in the above section, the same method can be applied even for the black holes.

  The metric of the BTZ black hole
is written as
\bal
ds^{2}&=-(r^{2}-R^{2})dt^{2}+\f{R^{2}}{r^{2}-R^{2}}dr^{2}+r^{2}d\phi^{2}\no
&=R^{2}(-\sinh^{2}\rho dt^{2}+d\rho^{2}+\cosh^{2}\rho
d\phi^{2})\no
&-\infty<t<\infty,\pi L<\phi<\pi L%/\beta
\label{btzmetric}
\eal
where the Euclidean time $\tau$ of BTZ black hole is periodic under $\tau\sim\tau+2\pi$ in order to make the Euclidean metric at $r=R$ smooth. For the BTZ black hole, the spacial direction is also periodic $\phi\sim\phi+2\pi L$ while for the Rindler-AdS it is non-compact $-\infty<\phi_r<\infty$. $L$ is proportional to the black hole temperature. The periodicities in the $\phi$ and $\tau$  directions reflect the fact that the Euclidean boundary of the three dimensional black hole is topologically equivalent to a torus, while that of the Rindler spacetime is equivalent to a two dimensional sphere.

\subsection{Thermofield double states}\label{TFD}
First we consider the bulk local state in the BTZ black holes dual to the thermofield double states \cite{Ete}. A thermofield double state is expressed as an entangled state between two non interacting CFTs: CFT$_1$ and CFT$_2$:
\ba
|\Psi_{\rm TFD}\lb\propto \sum_{E}e^{-\beta E/2}|E\lb_{1}\otimes|E\lb_{2},
\ea
where $\beta$ is a period of the Euclidean time which is $\beta=2\pi$ in our coordinate system.
Two CFTs are defined on a cylinder: $-\pi L<\phi<\pi L, -\infty<t<\infty$ while in the Rindler coordinate, CFTs are defined on a plane: $-\infty<\phi_r<\infty, -\infty<t<\infty$.
\begin{figure}[h!]
\begin{center}
  \includegraphics[width=9.5cm]{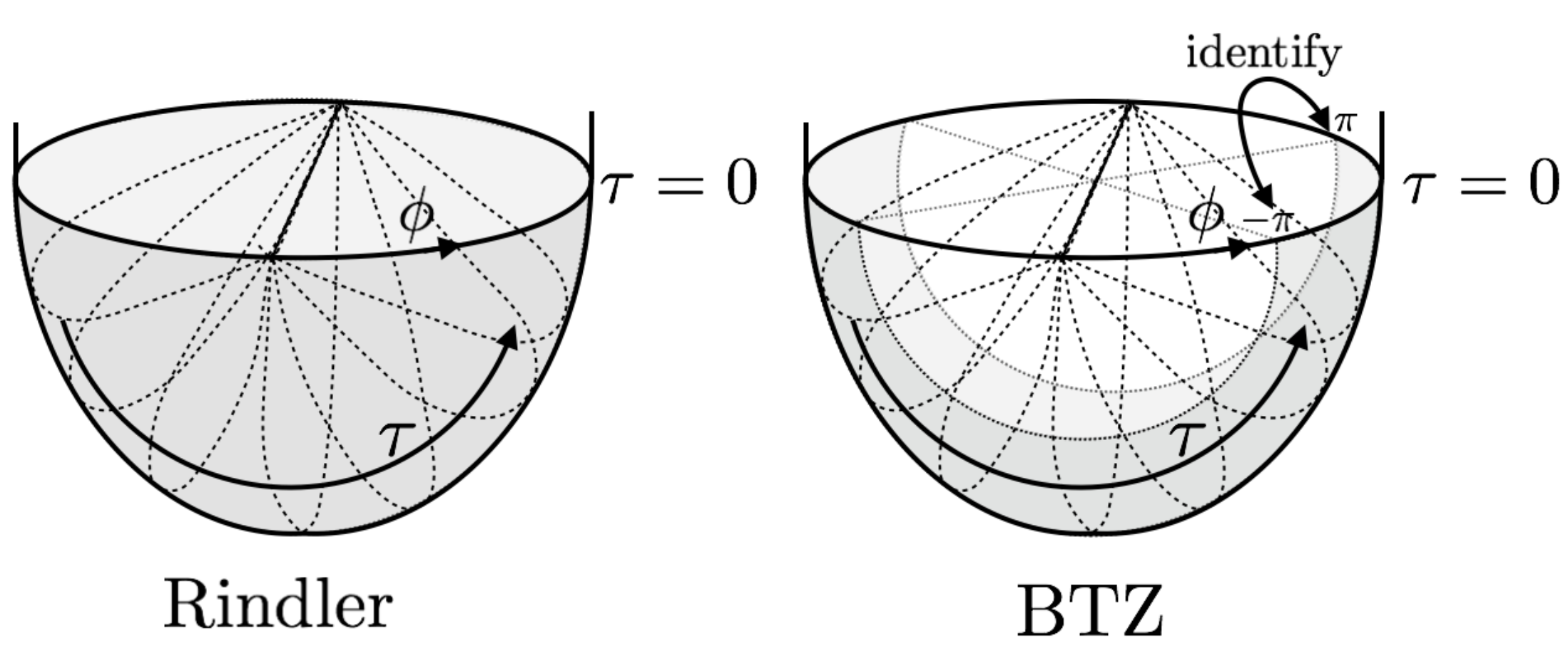}
  \caption{Euclidean boundaries of the Rindler-AdS and the BTZ black hole. BTZ black holes are locally equivalent with the Rindler-AdS coordinate and they can be constructed by identifying the spatial coordinate $\phi$ in the Rindler coordinate as $-\pi L<\phi<\pi L$.}
  \label{RinBTZ}
 \end{center}
\end{figure}
A thermofield double state is defined by the Euclidean path-integral over the half torus while the Rindler spacetime is over the half sphere (see Figure \ref{RinBTZ}). A subregion $-\pi L<\phi<\pi L$ in the Rindler boundary is equivalent to the boundary of the BTZ black holes. Thus the BTZ black hole spacetime and Rindler coordinate of the pure AdS spacetime is locally equivalent. This fact enables us to construct the bulk local states in the similar way as the Rindler coordinate.

For the construction of the bulk local states in the Rindler coordinate, we used the map between the boundary points $(\phi,\gamma)$ and the bulk points $(\rho,\phi,t)$ (\ref{Right})(\ref{Future}).
We assume that we can use the same map even for the BTZ black holes. We will see that calculations of the two point functions of the bulk local states in the BTZ which we will define below show that this assumption is indeed valid in the semi-classical limit. We introduce $(z,\z)$ coordinate locally around a point $(\phi,\gamma)$ on the torus as
\bal
\tanh(\f{\phi\pm t}{2})&=\tan (\f{\phi'\pm t'}{2})\no
z=e^{\tau'+i\phi'},&\ \z=e^{\tau'-i\phi'},\label{mapbtzz}
\eal
where $\tau'$ is the Euclidean time defined as $\tau'=it'$. Even though the map (\ref{mapbtzz}) is not defined globally since the sphere and the torus are topologically inequivalent, it is sufficient for our purpose of constructing bulk local states since the bulk local states in the Rindler-AdS (\ref{rightexpre})(\ref{futureexpre}) are locally expressed around $(\phi,\gamma)\LR(z_0,\z_0)$.%\footnote{The readers might wonder whether the Virasoro generators (\ref{Vira}) is locally defined or not since it uses the conformal transformation $f_{\boldsymbol{x}}(z)$  in the expression $f_{\boldsymbol{x}}^{*}T(z)$.However, once one writes it explicitly as $(\f{\pp z'}{\pp z})^{2}T(z')$ in the integral, one can see that it is locally expressed around $z=z_0$ (or equivalently $z'=0$). }.
We define Virasoro generators around the point $(\phi,\gamma)\LR(z_0,\z_0)$ similarly to the Rindler coordinate
\bal
L^{{\boldsymbol{x}}}_n\CO(z_0)&=\oint_{z_0} dz z^{n+1}\left(\f{\pp z'}{\pp z}\right)^2T(z')\CO(z_0),\no
\bar{L}^{{\boldsymbol{x}}}_n\CO(z_0)&=\oint_{\z_0} d\z \z^{n+1}\left(\f{\pp \z'}{\pp \z}\right)^2\bar{T}(\z')\CO(\z_0),\label{ViraBTZ}
\eal
where we can evaluate $\f{\pp z'}{\pp z}$ and $\f{\pp \z'}{\pp \z}$ through (\ref{Right}) for outside the horizon and (\ref{Future}) for inside the horizon. Based on the map (\ref{Right})(\ref{Future}) and the definitions of the Virasoro generators (\ref{ViraBTZ}), we propose the expressions for the bulk local states in the BTZ black holes dual to the thermofield double states as follows;
\bal
|\phi_\a(\rho,\phi,t)\lb_{\rm BTZ}&\equiv\sum_{k=0}^{\infty}(-1)^k\f{\Gamma(\Delta)}{k!\Gamma(k+\Delta)}
(L^{{\boldsymbol{x}}}_{-1})^{k}(\bar{L}^{{\boldsymbol{x}}}_{-1})^{k}\CO_\a(\phi,\gamma)|\Psi_{\rm TFD}\lb.\label{localbtz}
\eal
Since this definition of the bulk local states is given locally around the point $(\phi,\gamma)$, it is well defined on the torus.
\begin{figure}[h!]
\begin{center}
  \includegraphics[width=4.5cm]{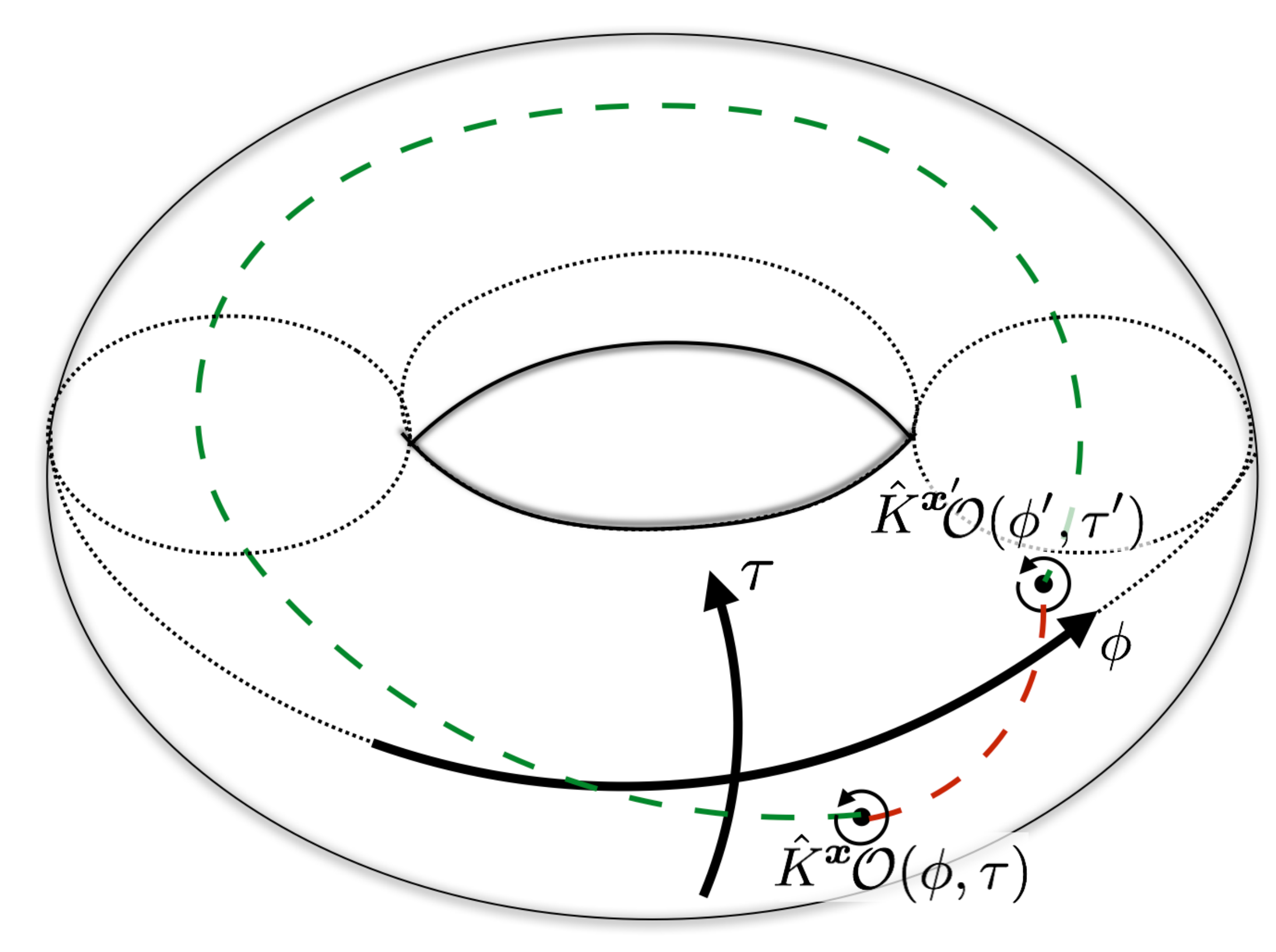}
  \caption{The bulk local state in BTZ black hole is defined as a CFT primary operator dressed by the localizing operator $\hat{K}^{\boldsymbol{x}}$ which is inserted at the point $(\phi,\gamma)$ on a half torus. Two point functions of the bulk local states are calculated on the torus. It can be expressed as a sum of correlators of mirror images. Each contribution comes from the direct path (red dashed line) or an incontractible winding path which goes around the circle along the $\phi$ direction (green dashed line).}
  \label{RinBTZ2}
 \end{center}
\end{figure}
\\ \\ \\ \\
\textbf{Correlators}\\
We calculate two point functions of the bulk local states defined on the thermofield double state dual to a BTZ black hole (\ref{localbtz}). The calculations are similar to the Rindler-AdS but we should take into account for an identification $\phi\sim\phi+2\pi L$ for BTZ black holes. In the semi-classical limit:$c\r \infty$, local operators in the large $c$ CFTs behave as free fields which are called generalized free fields. For such fields, we expect that correlators on a torus can be calculated by the method of mirror images especially in the high temperature limit. Correlators of mirror images are given by those in the Rindler coordinate, thus we have
\ba
\la \Psi_{\rm TFD}|\phi(\rho,\phi,t)\phi(\rho',\phi',t')|\Psi_{\rm TFD}\lb=\sum_{m,n}G^{\rm Rindler}(\rho,\phi+2\pi m,t;\rho',\phi'+2\pi n,t').
\ea
$G^{\rm Rindler}$ in the sum comes from the direct path or an incontractible winding path which goes around the circle along the $\phi$ direction.
This is the result expected from the bulk calculations \cite{IS}\cite{LO}\cite{Steif}.
\\ \\
\textbf{Black hole singularity}\\
We make a comment on the behavior of correlators when the bulk local states approach the center of the black hole:$r\r 0\LR \rho_f\r\pi/2$. In the Rindler construction, we saw that the generator of the $\rho_f$ direction is the same as the Hamiltonian in the global coordinate. If we evolve the bulk local state in the global time by $t_{\rm global}=\pi/2$, it becomes the Ishibashi state with no twist: $|J_\a\lb$,
\bal
\lim_{\rho_f\r\pi/2}|\phi_\a(\rho_f,\phi_f,t_f)\lb&= \lim_{\rho_f\r\pi/2} e^{-i(L_{1}+L_{-1}+\bar{L}_{1}+\bar{L}_{-1})t_{r}/2}e^{-i(L_{1}+L_{-1}-\bar{L}_{1}-\bar{L}_{-1})\phi_{r}/2}e^{i(L_{0}+\bar{L}_0)\rho_{f}}|\phi\lb\no
&=e^{-i(L_{1}+L_{-1}+\bar{L}_{1}+\bar{L}_{-1})t_{r}/2}|J_{\a}\lb,
\eal
which is invariant under the rotation of $\phi$. This implies that correlation functions in a BTZ black hole will be divergent if a bulk local operator approaches $\rho_f=\pi/2$ since mirror images give the same contributions to the correlator. Actually, we can see that the AdS-invariant distance $\sigma(\boldsymbol{x}_f|\boldsymbol{x})$ (\ref{sigma}) becomes independent of $\phi_f$ in the $\rho_r\r \pi/2$ limit. In this way, we can see the black hole singularity from the boundary CFT perspective.\newpage
\subsection{Single-sided black holes dual to the boundary states}\label{Bou}
\begin{figure}[h!]
\begin{center}
  \includegraphics[width=7.5cm]{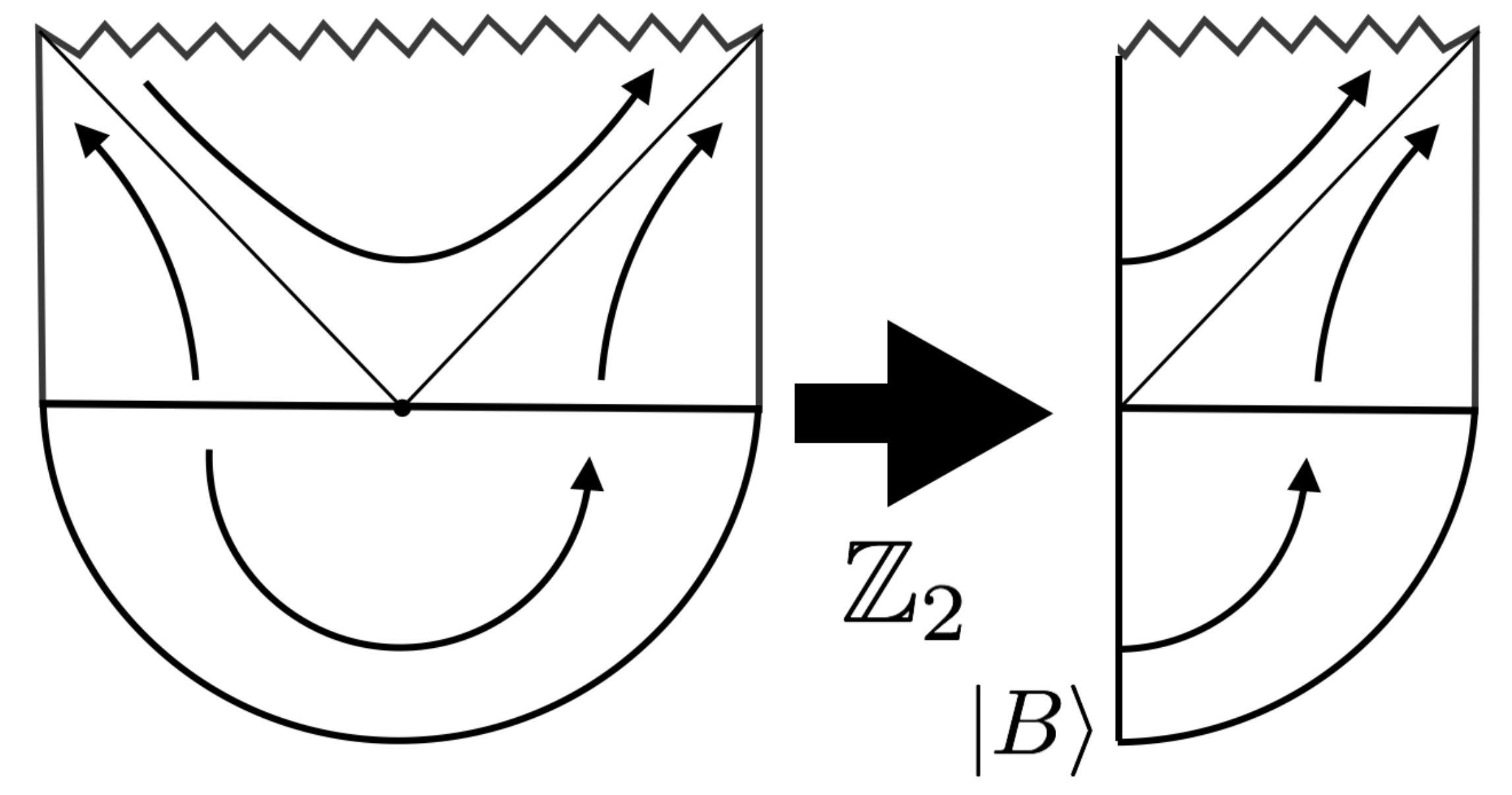}
  \caption{Single-sided black holes can be obtained by the $\mathbb{Z}_2$ identifications of two sided black holes. They are dual to the boundary states evolved by $\beta /4$ in the Euclidean time. $\beta$ is a temperature of the black hole.}
  \label{bounda}
 \end{center}
\end{figure}
We consider the bulk local states in the single-sided black holes obtained by the $\mathbb{Z}_2$ identifications of two sided black holes. This is dual to the boundary state evolved by $\beta/4$ in the Euclidean time\footnote{The different $\mathbb{Z}_2$ identification gives the so called geon geometry which is dual to the cross-cap states. The arguments for the bulk local states in the geon geometry are the same as the black holes dual to the boundary state.} and is interpreted as a holographic dual of a quantum quench \cite{HaMa}
\ba
|\Psi_{\rm BH}\lb=e^{-\beta H/4}|B\lb.\label{singlebh}
\ea
We expect that the same construction as the thermofield double state is valid for this type of the single-sided black holes. We define bulk local states in the black holes dual to (\ref{singlebh}) as
\bal
|\phi(\rho,\phi,t)\lb_{\Psi_{\rm BH}}&\equiv\sum_{k=0}^{\infty}(-1)^k\f{\Gamma(\Delta)}{k!\Gamma(k+\Delta)}
(L^{{\boldsymbol{x}}}_{-1})^{k}(\bar{L}^{{\boldsymbol{x}}}_{-1})^{k}\CO(\phi,\gamma)|\Psi_{\rm BH}\lb,\label{localbtzs}
\eal
where the map between the boundary and bulk points and the definition of the Virasoro generators are the same as those in the case of the thermofield double state. When we evaluate two point functions of the bulk local states, we also need to take a sum over mirror images introduced by the existence of the boundary $|B\lb$ in addition to the sum over the images from the identification of $\phi$. Away from the boundary, two point functions are evaluated as
\bal
\la\Psi_{\rm BH}|\phi(\rho,\phi,t)\phi(\rho',\phi',t')|\Psi_{\rm BH}\lb=\sum_{m,n}G^{\rm Rindler}(\rho,\phi+2\pi m,t;\rho',\phi'+2\pi n,t'),
\eal
since mirror images to $G^{\rm Rindler}$ introduced by the existence of the boundary give the same contributions to the correlators. Again we obtained the same results expected from the semi-classical calculations in the black hole backgrounds.
\subsection{Bulk local states from the heavy primary states}\label{Hea}
Finally, we attempt to construct the bulk local states in the geometry dual to a heavy primary state
\ba
|\Psi_{\rm BH}\lb=\CO_{H}(0)|0\lb,
\ea
whose conformal dimension is $\Delta_H=\CO(c)$. Especially we are interested in the case $\Delta_H=h+\bar{h}>c/12$. We focus on the spin-less primary states $h=\bar{h}$ for simplicity.

For light excited states in the large $c$ CFTs such like $\CO_{L_1}\ddd \CO_{L_n}|0\lb$, the correlators factorize into two point functions at the leading order of $1/c$ expansion. Thus we expect that we can apply the same definition of the bulk local states to the states $\CO_{L_1}\ddd \CO_{L_n}|0\lb$ as the vacuum state. This is related to the fact that the gravitational back reactions of the light fields $\CO_{L_i}$ is $\CO(\Delta_{L_i}/c)$ and as far as $\Delta_{L_i}\ll c$ and considering low point functions; $n\ll c$, we can neglect the modification of the background geometry from pure AdS and the arguments in the vacuum state is applicable.
However, if $\Delta_{H}= \CO(c)$, we cannot neglect the modifications of the geometry. The background $|\Psi_{\rm BH}\lb$ has the energy expectation value of order $c$,
\ba
\la T(\tilde{z})\lb_{\Psi_{\rm BH}}=\f{h_H}{\tilde{z}^2}\propto c
\ea
where $\tilde{z}$ is the complex plane coordinate defined as $\tilde{z}=e^{\tau+i\phi}$ and $(\tau,\phi)$ is a coordinate on the cylinder defined on the Euclidean boundary of AdS.
This implies the dual geometry is modified to a BTZ black hole whose temperature is $\beta=2\pi/|\a_H|, |\a_H|=\s{24h_H/c-1}$\cite{Bana}:
\bal
%ds^{2}&=-\f{r^{2}-r_0^2}{R^2}dt^{2}+\f{R^{2}}{r^{2}-r_0^{2}}dr^{2}+r^{2}d\phi^{2}\no
ds^{2}&=-(r^{2}-r_0^2)dt^{2}+\f{dr^{2}}{r^{2}-r_0^2}+r^{2}d\phi^{2}\no
&-\infty<t<\infty,-\pi<\phi<\pi,\label{btzmetric2}
 \eal
where $r_0=2\pi/\beta$ and the Euclidean time $\tau$ is periodic $\tau\sim \tau+\beta$\footnote{In this section, we set the AdS radius $R$ to be unity for simplicity. The metric (\ref{btzmetric2}) is equivalent to (\ref{btzmetric}) under the rescaling $r\r r_0 r,\phi\r 2\pi \phi/\beta, t\r 2\pi t/\beta$. }. Thus we expect that the construction of bulk local states for the vacuum state is not applicable for the heavy primary states whose dimensions are order $c$. In this section, we discuss how we should construct the bulk local states for such states.

Before considering bulk local states, we review the behavior of the correlation functions of light operators on the heavy primary states $|\Psi_{\rm BH}\lb$. In order to calculate the correlators on heavy primary states, it is useful to move to the uniformizing coordinate $w$ \cite{FKW2} where the expectation value of the energy momentum tensor becomes
\ba
\la T(w)\lb_{\Psi_{\rm BH}}=0.\label{tzero}
\ea
This coordinate can be obtained by the conformal transformation
\ba
w=\tilde{z}^{\a_H},\ \bar{w}=\bar{\tilde{z}}^{\bar{\a}_H}.\label{mapzw}
\ea
where $ \a_H=\s{1-\f{24h}{c}}$.
 \begin{figure}[h!]
\begin{center}
  \includegraphics[width=12cm]{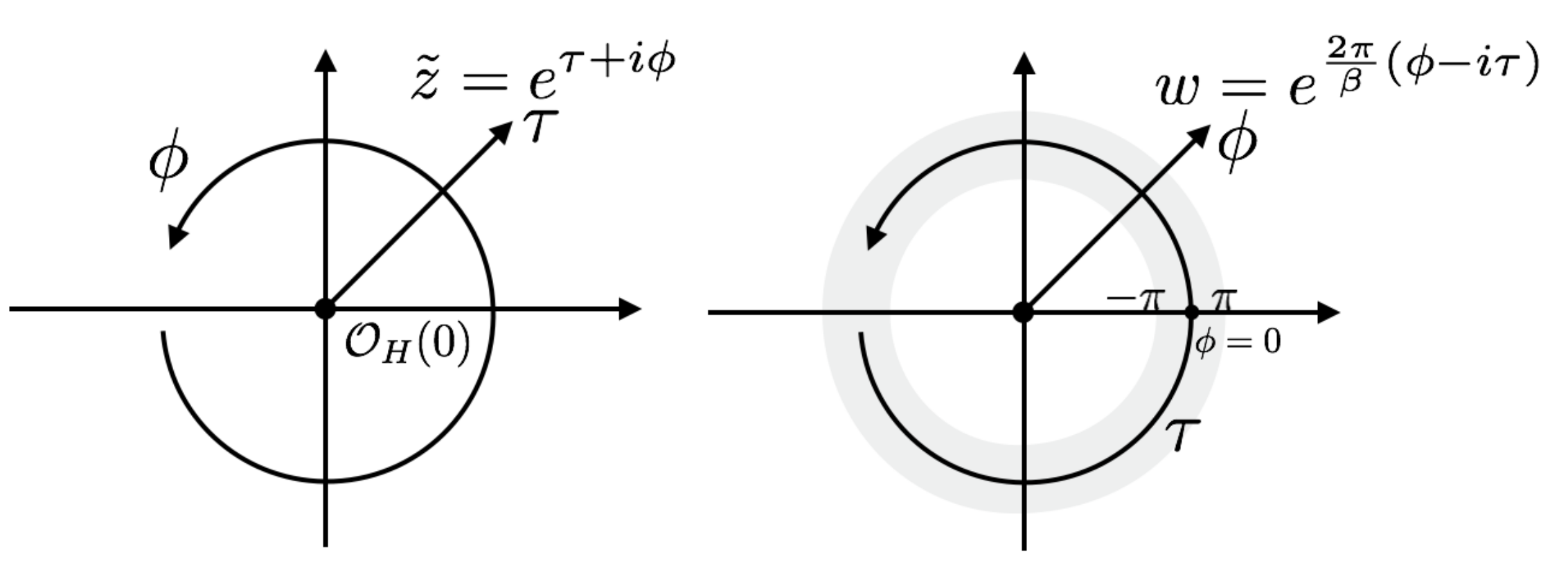}
  \caption{We move to the uniformizing coordinate $w$ by the conformal transformation where the effect of the insertions of the heavy operators $\CO_H$ is replaced by the background geometry. The Euclidean plane $(\tilde{z},\bar{\tilde{z}})$ is mapped to the gray region in the $w$ coordinate;$\phi_r$ and $\phi_r+4\pi^2/\beta$ are same point in the $\tilde{z}$ coordinate. It expresses the monodromy around the black hole singularity.}
  \label{heavyBTZ}
 \end{center}
\end{figure}
Non-tensor behavior of the CFT energy momentum tensor under conformal transformations
\ba
T(w)=\left(\f{d\tilde{z}}{dw}\right)^2T(\tilde{z}(w))+\f{c}{12}\{\tilde{z},w\}
\ea
enables us to move to a coordinate system which satisfies (\ref{tzero}) by the conformal map (\ref{mapzw}).
This coordinate system $w=e^{\f{2\pi}{\beta}(\phi-i\tau)}$ looks like the Rindler coordinate $\zeta=e^{\phi_r-i\tau_r}$ with a rescalling $\phi_r=2\pi\phi/\beta,\tau_r=2\pi\tau/\beta$. Notice that different points $(\phi_r,\tau_r)$ and $(\phi_r+4\pi^2/\beta,\tau_r)$ in the $w$ coordinate are the same point in the original coordinate $\tilde{z}$ (see Figure \ref{heavyBTZ}). This expresses the monodromy around the black hole singularity. 

Let us consider the four point function of the form $\la \CO_H\CO_\a(z) \CO_\a(z') \CO_H\lb$ where $z$ and $z'$ are points in the Euclidean coordinate $z$. The transformation (\ref{mapzw}) doesn't add any branch cuts to the four point functions at the leading order of $1/c$ expansion and the results are single-valued in the $w$ coordinate \cite{Hart}\cite{FKW1}\cite{FKW2}%\footnote{Of course, if we place primary operators in the Lorentzian time directions, we have branch cuts in the null directions which respect the causality, but this type of branch cuts is different from those created by the coordinate transformation (\ref{mapzw}).}
. Especially we consider the region where the vacuum conformal block in the s-channel $\CO_H\CO_H\r \CO_\a\CO_\a$ dominates the four point function in a holographic CFT. For such cases, we can compute the correlator as
\bal
\la \CO_H\CO_\a \CO_\a \CO_H\lb&=\sum_{\CO}\la \CO_H\CO_H \mathbb{P}_{\CO}\CO_\a \CO_\a \lb\sim\la \CO_H\CO_H \mathbb{P}_{0}\CO_\a \CO_\a \lb\no
&=\la \CO_H\CO_H\lb_w \la\CO_\a \CO_\a \lb_w+\CO(1/c)\no
&=\la\CO_\a \CO_\a \lb_w+\CO(1/c),\label{vacu}
\eal where
 $\mathbb{P}_{\CO}$ is a projector onto the irreducible representation of the Virasoro algebra with primary $\CO$ and we normalize $\la \CO_H\CO_H\lb_w$ to unity. In the second line, we used the fact that (\ref{tzero}) guarantees that the Virasoro descendant states in $\mathbb{P}_{\CO}$ such as $\f{1}{\s{\CN_{m_i,k_i}}}L^{k_1}_{-m_1}\ddd L^{k_n}_{-m_n}|0\lb$ can be neglected at the leading order of $1/c$ because the norm $\CN_{m_i,k_i}$ scales like $\CO(c)$ when $m_i>1$.
We can see from the above expressions that the two point functions for the heavy primary state $|\Psi_{\rm BH}\lb$ are reduced to two point functions for the vacuum state evaluated in the $w$ coordinate.
This manifests the thermality of the black holes; we can calculate correlators on $|\Psi_{\rm BH}\lb$ as if they are placed on a thermal background (or equivalently on a torus) as long as we only consider the leading order of $1/c$ expansion \cite{FKW2}.
 \begin{figure}[h!]
\begin{center}
  \includegraphics[width=5cm]{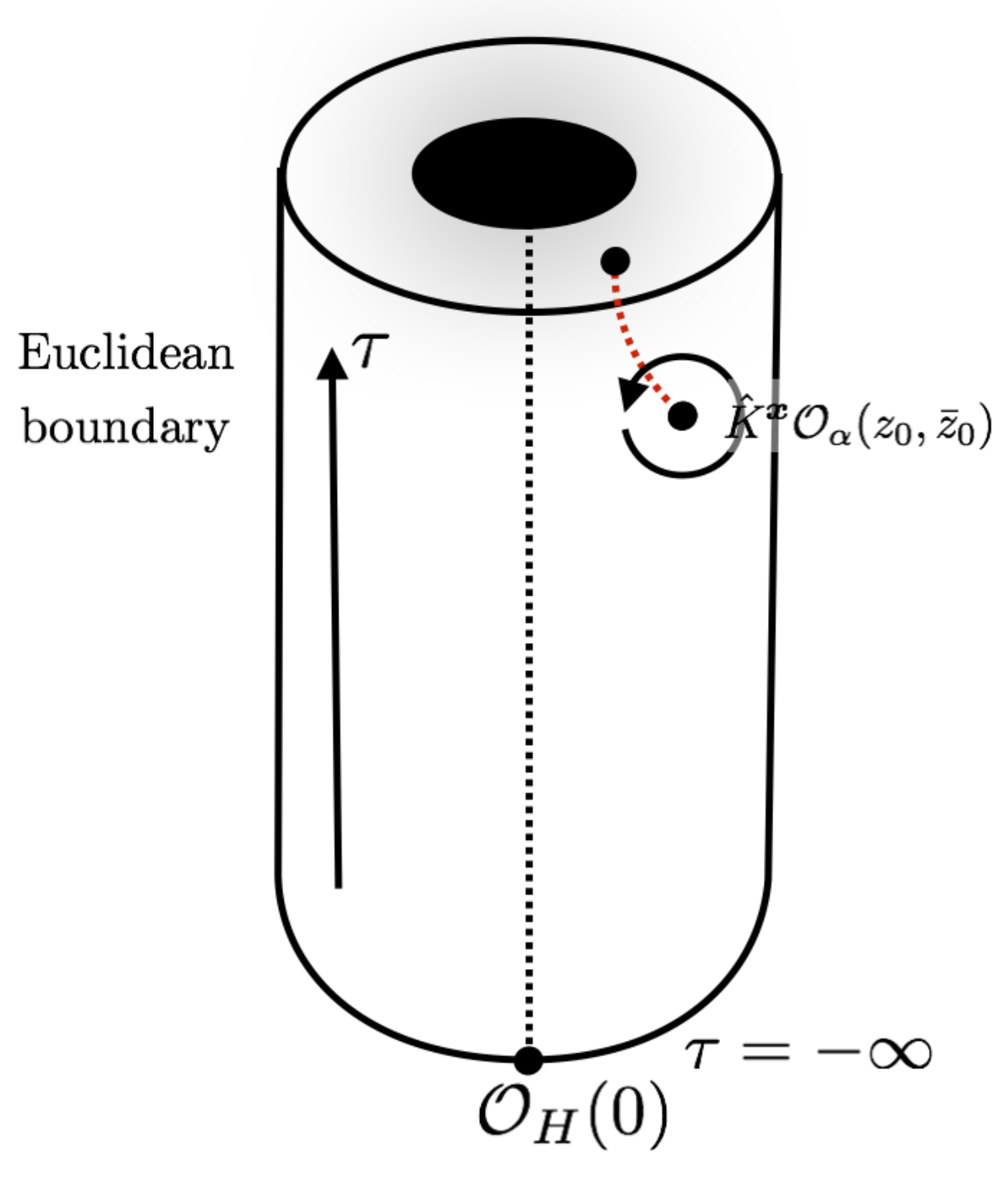}
  \caption{We can use the same way to construct the bulk local states from the heavy primary states as far as these background states look ``thermal''. Thermality of the correlators are brought  by the vacuum conformal block approximation of the correlators.}
  \label{heavybulklocal}
 \end{center}
\end{figure}

On the thermal background, we know how to construct bulk local states which reproduce the results expected from the semi-classical calculations in the gravity side as we saw in the previous section: the construction on the thermofield double state. Thus we propose the expression of the bulk local states on heavy primary states $|\Psi_{\rm BH}\lb$ as follows:
\bal
\hat{\phi}_\a(\rho,\phi,t)|\CO_H\lb&\equiv\sum_{k=0}^{\infty}(-1)^k\f{\Gamma(\Delta)}{k!\Gamma(k+\Delta)}
(L^{{\boldsymbol{x}}}_{-1})^{k}(\bar{L}^{{\boldsymbol{x}}}_{-1})^{k}\CO_\a(\phi,\gamma)|\CO_H\lb.
\eal
The map between the boundary points and the bulk points and the definition of the Virasoro generators $L^{{\boldsymbol{x}}}_{-1},\bar{L}^{{\boldsymbol{x}}}_{-1}$ are the same as the Rindler construction (\ref{Right})(\ref{RVira})(\ref{Future})(\ref{Vira}) with the identification $\phi_r=\f{2\pi}{\beta}\phi,\tau_r=\f{2\pi}{\beta}\tau$ where $(z,\z)$ coordinate is defined locally around the point $(\phi,\gamma)$ where the primary operator is inserted as
\bal
\tanh(\f{\pi}{\beta}(\phi\pm t))&=\tan (\f{\phi'\pm t'}{2})\no
z=e^{\tau'+i\phi'},&\ \z=e^{\tau'-i\phi'},
\eal
where we defined $\tau'=it'$.
As we saw, as far as the thermal background approximation, i.e, the approximation of the heavy-heavy-light-light correlators to its vacuum conformal block is valid, two point functions of the bulk local states defined above will reproduce the same result as the bulk local states constructed from the thermofield double states;
\bal
\la\CO_H|\phi(\rho,\phi,t)\phi(\rho',\phi',t')|\CO_H\lb&\sim \la\phi(\rho,\phi,t)\phi(\rho',\phi',t')\lb_{w}+\CO(1/c)\no
&=\sum_{m,n}G^{\rm Rindler}(\rho,\phi+2\pi m,t;\rho',\phi'+2\pi n,t')+\CO(1/c).\nonumber
\eal
In the second line the summation of mirror images is introduced since in the $w$ coordinate we have an identification $\phi\sim \phi+2\pi$ originated from the period of the spatial coordinate on the cylinder.\\ \\
\textbf{Discussions}\\
In this discussion section\footnote{After we complete this paper we got aware 
of the very
interesting paper \cite{PTK} on the net where a break down of semiclassical approximation and its connection to the absence of the other boundary of a BTZ black hole have been analyzed, which has a 
partial overlap with this discussion part of our paper.}, let us consider the following bulk-boundary correlator
\ba
 \la\CO_H|\phi_\a(\rho,\phi,0)\CO_\a(0,0)|\CO_H\lb.
 \ea for simplicity.
 This correlator can be calculated from the heavy-heavy-light-light correlator such as
 \ba
 \la\CO_H|\CO_\a(\phi,\tau)\CO_\a(0,0)|\CO_H\lb.\label{HHLL}
 \ea
 and correlators for descendants of $\CO_\a$. The boundary position $\tau$ corresponds to the bulk position $\rho$ via the relation $\tan \f{\pi}{\beta}(\tau-\beta/4)=\tanh\f{\rho}{2}$.

In our semi-classical calculations above, we assumed that the vacuum conformal block in the s-channel $\CO_H\CO_H\r \CO_\a\CO_\a$ with the identity sector intermediating dominates the heavy-heavy-light-light correlator. In the case where two light operators are close to each other, we expect that this assumption is indeed valid. This corresponds to the case where the bulk local operator is inserted near the boundary. If we move to the bulk interior, we should place the primary operator at $\tau<0$ and the distance between the light operators gets longer and the light and the heavy operator get closer. We expect that there is a phase transition somewhere during moving to the bulk interior, whe
re the semiclassical approximation breaks down, for which the t-channel $\CO_H\CO_\a\r \CO_H\CO_\a$ computations may be more useful. There is a possibility that it prevents us from constructing bulk local states which behave as if  they are  in ``the second region'' connected via a wormhole while we can construct them in the case of the thermofield double state. This phase transition will be also very important to know the spacetime structure near the horizon and black hole interior while we need to analyze the saddle points of the Lorentzian correlators.
In order to know the point where the transition occurs exactly, we need more details of the CFT.

\newpage
\section{Conclusions}
%%%%%%%%%%%%%%%%%%%%%%
%%%%%%%%%%%%%%%%%%%%%%

In this paper we studied the CFT duals of locally excited states in bulk spacetimes (we simply call them bulk local states) in various setups of AdS/CFT correspondence. By employing the conformal symmetry, we gave a systematical construction of the locally excited states in the Poincare AdS
(appendix A), AdS Rindler (Section 2.3), two sided BTZ black hole (section 3.1) and 
single-sided black hole dual to a pure state (section 3.2 and 3.3). Though we focused on a two dimensional CFT dual to a three dimensional bulk space, part of our results (i.e. Poincare AdS and AdS Rindler) can be generalized to those in higher dimensions straightforwardly.

Our constructions are mainly divided into two procedures. The first step is to find a map between the boundary coordinate where a holographic CFT lives and those of the bulk AdS. This is obtained by considering geodesics which connect between boundary points and bulk points as we confirmed from the viewpoint of $SL(2,R)$ conformal symmetry. Accordingly, to find a CFT dual of a locally excited state at a bulk point, we start with its dual primary operator in the CFT at the boundary point by applying the mentioned map to the bulk point.  The second step is to dress the primary operator by acting global Virasoro operators in an appropriate way. These complete our construction.

Our construction method is background independent since we can use the same expressions both for the Rindler-AdS and BTZ black holes. The similar arguments have been discussed in the context of the HKLL construction of bulk local operators in the three dimensional AdS spacetime \cite{HKLL2}. As we saw in subsection \ref{Hea}, our construction is also independent of the CFT states as far as their dual geometries look thermal.

We expect that our bulk local state construction is useful to understand interior structures of black holes in AdS/CFT. Indeed in this paper we probed the interior of single-sided black hole created by a heavy primary operator by using our bulk local states. We find an implication that the large central charge approximation breaks down if we try to move from one boundary to the other due to the phase transition between $s$-channel and $t$-channel. This is consistent with the expectation that it is a single-sided black hole as it is dual to a pure state. It will be a very interesting future problem to examine this phase transition quantitatively.
\\ \\ 
 {\bf Acknowledgements}

% We thank *** for useful conversations. 
We are very grateful to Kento Watanabe for collaborations at an early stage. KG is supported by JSPS fellowships. TT is supported by the Simons Foundation through the ``It from Qubit'' collaboration. TT is supported by JSPS Grant-in-Aid for Scientific Research (A) No.16H02182. TT is also supported by World Premier International Research Center Initiative (WPI Initiative) from the Japan Ministry of Education, Culture, Sports, Science and Technology (MEXT). KG and TT thank very much the long term workshop ``Quantum Information in String Theory and Many-body Systems'' held at YITP, Kyoto where this work was initiated.

\newpage
\appendix
\section{Representation of $SL(2,\mathbb{R})$}
\subsection{Embedding formalism of AdS}
AdS$_3$ is defined as the hyperboloid $-X_0^2-X_1^2+X_2^2+X_3^2=-1$\footnote{In this section we set $R=1$ for simplicity.} embedded in a space with metric $ds^2= -dX_0^2-dX_1^2+dX_2^2+dX_3^2$. The global coordinate of AdS can be embedded as follows.
\begin{alignat}{2}
X^0&=\cosh\rho \cos t\no
X^1&=\cosh\rho \sin t\no
X^2&=\sinh\rho \sin \phi\no
X^3&=\sinh\rho \cos \phi
\end{alignat}
The metric in the global coordinate is written as
\bal
ds^{2}&=-\cosh^{2}\rho dt^{2}+d\rho^{2}+\sinh^{2}\rho
d\phi^{2}.
\eal
We introduce the coordinate in the Euclidean boundary as
\ba
z=e^{\tau+i\phi},\ \z=e^{\tau-i\phi}.
\ea
The Poincare coordinate is embedded as follows
\bal
X^0&=\f{1}{2y}(1+x^2+y^2-t'^2)\no
X^1&=\f{t'}{y}\no
X^2&=\f{x}{y}\no
X^3&=\f{1}{2y}(1-x^2-y^2+t'^2).
\eal
The metric of the Poincare coordinate is written as
\ba
ds^{2}=\f{-dt'^{2}+dy^{2}+dx^{2}}{y^{2}}.
\ea
We introduce the coordinate in the Euclidean boundary of the Poincare coordinate as
\ba
\xi=\tau'+ix,\ \bar{\xi}=\tau'-ix.
\ea
The right/left Rindler coordinates are embedded as follows
\begin{alignat}{2}
X^0&=r\cosh\phi_r&=&-r\cosh\phi_{l}\no
X^1&=\sqrt{r^2-1}\sinh t_r&=&\sqrt{r^2-1}\sinh t_l\no
X^2&=r\sinh \phi_r&=&-r\sinh \phi_{l}\no
X^3&=\sqrt{r^2-1}\cosh t_r&=&-\sqrt{r^2-1}\cosh t_{l}.
\end{alignat}
The metric in each wedge is written as
\bal
ds^{2}&=-(r^{2}-1)dt^{2}+\f{dr^{2}}{r^{2}-1}+r^{2}d\phi^{2}\no
&=-\sinh^{2}\rho dt^{2}+d\rho^{2}+\cosh^{2}\rho
d\phi^{2}\no
&r>1\ (\rho>0),-\infty<t<\infty,-\infty<\phi<\infty\eal
where we set $r=\sinh\rho$.
The left wedge and the right wedge are related by the analytic continuation $t_{l}=-t_{r}+i\pi$, $\phi_{l}=-\phi_{r}$. We introduce the coordinate in the Euclidean boundary of the Rindler coordinate as
\ba
\zeta=e^{\phi_r-i\tau_r},\ \bar{\zeta}=e^{\phi_r+i\tau_r}.
\ea
In the boundary limit, these coordinates are related as follows:
\ba
(x\pm t')/2=\tan (\phi\pm t)/2=\tanh (\phi_r\pm t_r)/2.
\ea
\subsection{$SL(2,\mathbb{R})$ symmetry}
The isometry of AdS$_3$ spacetime $SO(2,2)$ is realized as the rotations and boosts in the ambient space $\mathbb{R}^{2,2}$ \cite{KV}\cite{BKL}.
The generators of the $SO(2,2)$ consist of the rotation generators
\ba
L_{ab}=X^a\pp_b-X^b\pp_a
\ea
in the $ab=01,23$ planes, and the boost generators
\ba
J_{ab}=X^a\pp_b+X^b\pp_a
\ea
in the $ab=02,03,12,13$ planes.\\
The $SL(2,\mathbb{R})_L$ generators are the linear combinations
\ba
J_0=(L_{01}-L_{23})/2, \q J_1=(J_{02}+J_{13})/2,\q J_2=(J_{12}-J_{03})/2
\ea
and the $SL(2,\mathbb{R})_R$ generators are expressed as
\ba
\bar{J}_0=(L_{01}+L_{23})/2, \q \bar{J}_1=
(-J_{02}+J_{13})/2,\q \bar{J}_2=-(J_{12}+J_{03})/2.
\ea
They satisfy the commutation relations
\ba
[J_0,J_2]=J_1,\q [J_0,J_1]=-J_2,\q [J_1,J_2]=J_0\label{commuj}
\ea
and similarly for the $\bar{J}$.
\subsection{Bulk local operators}
We want to rconsider a scalar field $\hat{\phi}_\a$ in AdS dual to the spinless primary operator
$\CO_\a$ whose conformal dimension  $\Delta_\a=h_\a+\bar{h}_\a=2h_\a$ is related to the mass of the scalar field as $
\Delta_\a=1+\s{m^2_\a R^2+1}
$. Consider the bulk scalar fields at the center of the AdS; $\vec{X}_0=(X^0,X^1,X^2,X^3)=(1,0,0,0)$ which corresponds to the bulk point $(\rho,\phi,t)=(0,0,0)$ in the global coordinate, $(y,x,t')=(1,0,0)$ in the Poincare coordinate and $(\rho,\phi_r,t_r)=(0,0,0)$ in the Rindler coordinate.
 They should be invariant under the rotation/boosts around the point $\vec{X}_0$, thus they satisfy
 \ba
[L_{23},\hat{\phi}_\a(\vec{X}_0)]=[J_{12},\hat{\phi}_\a(\vec{X}_0)]=[J_{13},\hat{\phi}_\a(\vec{X}_0)]=0.\label{bls}
\ea
We will see explicit forms of the bulk local operators in the following sections.
\subsection{Representation in the elliptic basis}
First we consider the representation in the basis diagonalizing $J_0$ which is a compact direction of $SL(2,\mathbb{R})$:
\bal
L_0&=iJ_0,\q L_{\pm 1}= i(J_1\pm i J_2),\no
\bar{L}_0&=i\bar{J}_0,\q \bar{L}_{\pm 1}=i(\bar{J}_1 \pm i\bar{J}_2).
\eal
They satisfy the following commutation relations
\ba
[L_0,L_{\pm 1}]=\mp L_{\pm 1},\q [L_{+1},L_{-1}]=2L_0.
\ea
Notice that the $L_0$ is a Hermitian operator and $L_{\pm 1}$ are adjoint operators.\\
This basis is natural for the (radial) quantization in the global coordinate since it diagonalizes the Hamiltonian in the global coordinate. The elliptic basis can be written by the Killing vectors in the global AdS as follows
\bal
L_0&=i\pp_+\no
L_{\pm1}&=ie^{\pm ix^{+}}\biggl[\f{\cosh2\rho}{\sinh2\rho}\pp_+-\f{1}{\sinh2\rho}\pp_-\mp\f{i}{2}\pp_\rho\biggl]\no
\bar{L}_0&=i\pp_-\no
\bar{L}_{+1}&=ie^{\pm ix^{-}}\biggl[\f{\cosh2\rho}{\sinh2\rho}\pp_--\f{1}{\sinh2\rho}\pp_+\mp\f{i}{2}\pp_\rho\biggl]
\eal
where $x^\pm=t\pm x$.\\
The boundary limit of these basis becomes
\bal
L_n=-z^{n+1}\pp_z,\q
\bar{L}_n=-\z^{n+1}\pp_{\z}
\eal
for $n=0,\pm 1$ where $z=e^{ix^+}=e^{\tau+i\phi},\bar{z}=e^{\tau-i\phi}$ are coordinates in the Euclidean plane defined on the boundary of global AdS. This coordinate system cuts the complex plane as $\mathbb{S}\times \mathbb{R}$, and states in the global coordinate can be obtained by the Euclidean path-integrals from the origin to the unit circle. The $SL(2,\mathbb{R})$ generators are defined as Laurent coefficients of the energy-momentum tensor
\ba
L_n=\oint _{\mathcal{C}}\f{dz}{2\pi i}z^{n+1}T(z),\q
\bar{L}_n=\oint _{\mathcal{C}}\f{d\z}{2\pi i}\z^{n+1}\bar{T}(\z)
\ea
where $\mathcal{C}$ is a unit circle.
  Time inversion involves the inversion $|z|\lr 1/|z|$ thus
  \ba
  L_{\pm 1}^{\dg}=L_{\mp 1}.
  \ea
A primary state with respect to the elliptic basis satisfies
\bal
L_1|\CO_\a\lb&=\bar{L}_1|\CO_\a\lb=0,\no L_0|\CO_\a\lb=h|\CO_\a\lb,&\q \bar{L}_0|\CO_\a\lb=\bar{h}|\CO_\a\lb.
\eal
This state can be obtained by inserting a primary operator at the origin of the Euclidean coordinate $(z,\z)=(0,0)$ which is  $\tau=-\infty$ in the global coordinate 
\ba
|\CO_\a\lb=\CO_\a(z=\z=0)|0\lb.
\ea
\\
\textbf{Bulk local states in the elliptic basis}\\
We can write bulk local operators (\ref{bls}) explicitly in this elliptic basis. For simplicity we consider the bulk local states defined as states locally excited by the bulk local operator instead of the bulk local operators themselves;
\ba
|\phi_\a\lb=\hat{\phi}_\a(\vec{X}_0)|0\lb,
\ea
where $|0\lb$ is the $SL(2,\mathbb{R})$ invariant vacuum which corresponds to the pure (empty) AdS spacetime. The conditions for the bulk local operators can be written in the elliptic basis as
\ba
(L_0-\bar{L}_0)|\phi_\a\lb=(L_{\pm 1}+\bar{L}_{\mp 1})|\phi_\a\lb=0.
\ea
We can construct solutions for these equations as twisted Ishibashi states with respect to the elliptic basis of $SL(2,\mathbb{R})$ algebra
\ba
|\phi_\a\lb=\sum_{k=0}^{\infty}\f{\Gamma(\Delta)}{k!\Gamma(k+\Delta)}%(-1)^k
(-1)^k(L_{-1})^{k}(\bar{L}_{-1})^{k}|\CO_{\a}\lb,
\ea
where $|\CO_{\a}\lb$ is the state obtained by inserting the primary operator at $\tau=-\infty$ in the global coordinate or equivalently $\tau'=-1$ in the Poincare coordinate.
\ba
|\CO_\a\lb\equiv\CO_\a(\tau=-\infty)|0\lb=e^{-H_P}\CO_\a(x=t'=0)|0\lb.
\ea
In the AdS/CFT, this state is dual to a massive particle which is situated at the center of global AdS$_3$. It is equivalent to a massive particle moving in the Poincare AdS$_3$. This indeed perfectly agrees with the realization of locally excited state by a massive particle which is used to calculate the entanglement entropy under the local quenches \cite{NNT}.

The bulk local states at different bulk points are related by the $SL(2,\mathbb{R})$ transformations as follows:
\ba
|\phi_\a(\boldsymbol{x})\lb=g(\boldsymbol{x})|\phi_\a\lb.
\ea
where $g(\boldsymbol{x})$ is expressed as
\bal
{\rm global;}\q g(\rho,\phi,t)&=e^{i(L_0+\bar{L}_0)t}e^{\mathchar`-
\rho(e^{\mathchar`-i\phi}\frac{L_{1}-\bar{L}_{\mathchar`-1}}{2}+e^{i\phi}\frac{\bar{L}_{1}-L_{\mathchar`-1}}{2})}\no
{\rm Poincare;}\q g(y,x,t)
&=e^{i(L_0+\f{L_{-1}+L_{-1}}{2})x^+}e^{i(\bar{L}_0+\f{\bar{L}_{-1}+\bar{L}_{-1}}{2})x^-}y^{
\frac{L_{1}-\bar{L}_{\mathchar`-1}+\bar{L}_{1}-L_{\mathchar`-1}}{2}}.
\eal

\subsection{Representation in the hyperbolic basis (I)}
Next we consider the representation in the basis diagonalizing $J_2$ which is a non-compact direction of $SL(2,\mathbb{R})$:
\bal
L^h_0&=-J_2,\q L^h_{\pm 1}=i(J_1\mp J_0),\no
\bar{L}^h_0&=-\bar{J}_2,\q \bar{L}^h_{\pm 1}= i(\bar{J}_1 \mp \bar{J}_0).
\eal
The representation in the basis diagonalizing the non-compact directions make the spectrum continuous.
From (\ref{commuj}), they satisfy the following commutation relations
\ba
[L^h_0,L^h_{\pm 1}]=\mp L^h_{\pm 1},\q [L^h_{+1},L^h_{-1}]=2L^h_0.
\ea
Notice that the $L^h_0$ is a anti-Hermitian operator and $L^h_{\pm 1}$ are Hermitian operators. Anti-Hermitian operators in continuous spectrum can have real eigenvalues. The explicit relation between the elliptic basis and the hyperbolic basis is
\begin{alignat}{2}
L_0^{h}&=\f{L_1-L_{-1}}{2},&\bar{L}^h_0&=\f{\bar{L}_1-\bar{L}_{-1}}{2}\no
L_{\pm 1}^{h}&=\mp L_0+ \f{L_1+L_{-1}}{2}, &\bar{L}_{\pm 1}^{h}&=\mp\bar{L}_0+ \f{\bar{L}_1+\bar{L}_{-1}}{2}\no
L_0&=-\f{L^h_{1}-L^h_{-1}}{2},&\bar{L}_0&=-\f{\bar{L}^h_{1}-\bar{L}^h_{-1}}{2},\no
L_{\pm 1}&=\pm L_0^h+\f{L^h_{1}+L^h_{-1}}{2},
&\bar{L}_{\pm 1}&=\pm \bar{L}_0^h+\f{\bar{L}^h_{1}+\bar{L}^h_{-1}}{2}.\label{eh}
\end{alignat}
These two bases are related by the non-unitary transformation $U_h^e$: 
\bal
U_h^eL^h_n(U^e_h)^{-1}&=L_n\no
U^e_h\bar{L}^h_n(U^e_h)^{-1}&=\bar{L}_n\qq
U^e_h=e^{\f{\pi}{4}[L^{h}_1+L^{h}_{-1}+\bar{L}^{h}_1+\bar{L}^{h}_{-1}]}\no
U_e^hL_n(U_e^h)^{-1}&=L^h_n\no
U_e^h\bar{L}_n(U_e^h)^{-1}&=\bar{L}^h_n
\qq
U_e^h=e^{-\f{\pi}{4}[L_1+L_{-1}+\bar{L}_1+\bar{L}_{-1}]}.\label{ooo}
\eal
\\
\textbf{Poincare coordinate and hyperbolic representation}\\
In the Poincare coordinate, basis for $SL(2,\mathbb{R})$ diagonalizing the dilation operator corresponds the hyperbolic basis and one diagonalizing the Hamiltonian corresponds the parabolic basis as we will see.
The hyperbolic basis can be written by the Killing vectors in the Poincare AdS as follows
\bal
L_0^{h}&=-\f{y}{2}\pp_y-x^+\pp_+ %\r -x^+\pp_+
\no
L_{-1}^{h}&=
i\pp_+\no
L_{+1}^{h}&=
-ix^+y\pp_y-i(x^+)^2\pp_+-iy^2\pp_-%\r %i-i(x^+)^2\pp_+
\no
\bar{L}_0^{h}&=\f{-y}{2}\pp_y-x^-\pp_- %\r -x^-\pp_-
\no
\bar{L}_{-1}^{h}&=
i\pp_-\no
\bar{L}_{+1}^{h}&=-ix^-y\pp_y-i(x^-)^2\pp_--iy^2\pp_+.%\r %i-i(x^-)^2\pp_-.
\eal
where $x^{\pm}=t'\pm x$ and $\pp_{\pm}=(\pp_{t'}\pm \pp_x)/2$.
The boundary limit of these basis become
\bal
L^h_n=-\xi^{n+1}\pp_\xi,\q
\bar{L}^h_n=-\bar{\xi}^{n+1}\pp_{\bar{\xi}}
\eal
for $n=0,\pm 1$ where $\xi=\tau+ix ,\bar{\xi}=\tau-ix$ is the coordinates in the Euclidean boundary of the Poincare coordinate. This coordinate system cuts the complex plane as $\mathbb{R}\times \mathbb{R}$, and the states in the Poincare coordinate can be obtained by the Euclidean path integral from $\tau=-\infty$ to $\tau=0$. The $SL(2,\mathbb{R})$ generators are defined as Laurent coefficients of the energy-momentum tensor
\ba
L^h_n=\int_{-i\infty}^{i\infty}\f{d\xi}{2\pi i}\xi^{n+1}T(\xi),\q
\bar{L}^h_n=\int_{-i\infty}^{i\infty}\f{d\bar{\xi}}{2\pi i}\bar{\xi}^{n+1}\bar{T}(\bar{\xi}).
\ea
  Time inversion involves the complex conjugation $\xi\lr -\bar{\xi}$ thus
\ba
 L_{\pm 1}^{h\dg}=L^h_{\pm 1}.
\ea
Notice that the $L^h_0+\bar{L}^h_0$ corresponds the dilatation operator in the Euclidean plane and the Hamiltonian in the Poincare coordinate in the hyperbolic basis can be written as
\ba
H^P=L^h_{-1}+\bar{L}^h_{-1},
\ea
which is not diagonalized in this hyperbolic basis. The parabolic basis which diagonalizes $J_0- J_1$ can diagonalize the Poincare Hamiltonian. Thus one particle states in the Poincare coordinate $|\w,k\lb=a^{\dg P}_{\w,k}|0\lb_P$ transform according to the parabolic representation rather than the hyperbolic representation.\\
The relationship between the elliptic basis and hyperbolic basis can be obtained by the coordinate transformation between the global coordinate and the Poincare coordinate
\bal
(x\pm t')/2&=\tan (\phi\pm t)/2
 \LR z=f(\xi)=\f{1+\xi}{1-\xi}, \z=\bar{f}(\bar{\xi})=\f{1+\bar{\xi}}{1-\bar{\xi}}.
 \eal
 By using the conformal transformation $z=f(z'),\z=\bar{f}(\z')$, we can express $L^h_n$ and $\bar{L}^h_n$ as
 \bal
L^{h}_{n}&=\oint dz' z'^{n+1}f^{*}T(z')=\oint dz\f{(z+1)^{2}}{2}\biggl(\f{z-1}{z+1}\biggl)^{n+1}T(z)\no
\bar{L}^{h}_{n}&=\oint d\bar{z}' \bar{z}'^{n+1}\bar{f}^{*}\bar{T}(\bar{z}')=\oint d\z\f{(\z+1)^{2}}{2}\biggl(\f{\z-1}{\z+1}\biggl)^{n+1}\bar{T}(\z)
\eal
which reproduce the relation (\ref{eh}).
A primary state with respect to the hyperbolic basis satisfies the following equations
\bal
L^h_1|\CO_\a\lb&=\bar{L}^h_1|\CO_\a\lb=0,\no L^h_0|\CO_\a\lb=h|\CO_\a\lb,&\q \bar{L}^h_0|\CO_\a\lb=\bar{h}|\CO_\a\lb.
\eal
This state can be obtained by inserting a primary operator at the origin of the Euclidean boundary $(x,t')=(0,0)$ of the Poincare coordinate
\ba
|\CO^h_\a\lb=\CO_\a(x=t'=0)|0\lb,
\ea
which is a locally exited state in the Poincare coordinate.
These can be obtained from the primary states with respect to the generators in the elliptic basis by the non-unitary transformation $U^h_e$
\ba
|\CO^h_\a\lb=U^h_e|\CO_\a\lb.
\ea
\\
\textbf{Bulk local states in the hyperbolic basis}\\
We can write bulk local operators (\ref{bls}) explicitly in this  basis. The conditions for the bulk local states can be written in this basis as
\ba
(L^h_0-\bar{L}^h_0)|\phi_\a\lb=(L^h_{\pm 1}+\bar{L}^h_{\mp 1})|\phi_\a\lb=0.
\ea
This is just the same condition as that in the elliptic basis. Actually the bulk local states are invariant under the conformal transformation between the elliptic basis and hyperbolic basis
\ba
U^h_e|\phi_\a\lb=|\phi_\a\lb.
\ea
The bulk local states can be expressed as twisted Ishibashi states with respect to the hyperbolic basis of $SL(2,\mathbb{R})$ algebra.
\ba
|\phi_\a\lb=\sum_{k=0}^{\infty}\f{\Gamma(\Delta)}{k!\Gamma(k+\Delta)}%(-1)^k
(-1)^k(L^h_{-1})^{k}(\bar{L}^h_{-1})^{k}\CO_{\a}(\vec{x}=0)|0\lb.
\ea
where $\CO_{\a}(\vec{x}=0)|0\lb$ is the state obtained by inserting a primary operator at $\vec{x}=(x,t')=(\phi,t)=(0,0)$ in the Poincare/global coordinate.\\
The bulk local states at different bulk points are related by the $SL(2,\mathbb{R})$ transformations
\ba
|\phi_\a(\boldsymbol{x})\lb=g(\boldsymbol{x})|\phi_\a\lb
\ea
where $g(\boldsymbol{x})$ is given by
\bal
{\rm global;}\q g(\rho,\phi,t)
&=e^{-i\f{L^h_1-L^h_{-1}}{2}x^{+}}e^{-i\f{\bar{L}^h_{1}-\bar{L}^h_{-1}}{2}x^{-}}e^{-\rho\f{L^{h}_0+\bar{L}^{h}_0}{2}}\no
{\rm Poincare;}\q g(y,x,t)&=e^{i(L_{-1}^{h}+\bar{L}_{-1}^{h})t}e^{i(L_{-1}^{h}-\bar{L}_{-1}^{h})x}y^{L^{h}_0+\bar{L}^{h}_0}
\eal
Expressed in this basis, we can easily see the bulk local states satisfy the extrapolate dictionary of AdS/CFT;
\ba
\lim_{\rho\r \infty}e^{\rho\Delta}|\phi_\a(\rho,\vec{x})\lb=\lim_{y\r 0}y^{-\Delta}|\phi_\a(y,\vec{x})\lb=\CO_\a(\vec{x})|0\lb\label{extra}
\ea
We can easily  compute two point functions in the Poincare coordinate as follows. The boundary two point functions are expressed as
\ba
\la \CO(x,t)\CO(0)\lb=\f{1}{(x^2-t^2)^{\Delta}},
\ea
where the branch cut at $x\leq |t|$ respects the boundary causality.
\bal
\la \phi(y,x,t)\CO(0)\lb&=\sum_{k=0}^{\infty}y^{\Delta+2k}\f{\Gamma(\Delta)}{k!\Gamma(k+\Delta)}\pp_+^{k}\pp_-^k\la \CO(x,t)\CO(0)\lb\no
&=\sum_{k=0}^{\infty}(-1)^k\f{\Gamma(k+\Delta)}{k!\Gamma(\Delta)} y^{\Delta+2k}\f{1}{(x^2-t^2)^{\Delta+k}}\no
&=\f{y^{\Delta}}{(y^2+x^2-t^2)^{\Delta}}
\eal
This result is just the same as the bulk-to-boundary propagator in the Poincare coordinate and again branch cut at $\s{x^2+y^2}\leq |t|$ guarantees the causality in the bulk; $[\phi(y,\vec{x}),\CO(\vec{x}')]\neq 0$ when $(y,\vec{x})$ and $(0,\vec{x}')$ are time-like separated.\\ \\
\textbf{Equivalence with the HKLL construction in the Poincare coordinate}\\
In the Poincare coordinate, the bulk local states at $(y,x,t)$ can be written as follows.
\ba
|\phi_\a(y,x,t)\lb=\sum_{k=0}^{\infty}(-1)^ky^{\Delta+2k}\f{\Gamma(\Delta)}{k!\Gamma(k+\Delta)}%(-1)^k
(L^h_{-1})^{k}(\bar{L}^h_{-1})^{k}\CO_\a(x,t)|0\lb\label{ours}
\ea
The CFT representations of the bulk local fields can also obtained as smeared boundary primary operators which is known as HKLL construction \cite{HKLL0}\cite{HKLL1}\cite{HKLL2}. They are schematically expressed as
\bal
\phi_\a^{\rm HKLL}(\boldsymbol{x})&=\int dt'dx'^{d-1}K(\boldsymbol{x};x',t')\CO_\a(x',t'),
\eal
for an appropriate choice of the kernel $K(\boldsymbol{x};x',t')$.\\
In the Poincare coordinate, they can be written as follows \cite{HKLL2}
\bal
\phi_\a^{\rm HKLL}(y,x,t)&=\f{\Delta-1}{\pi}\int_{x'^2+t'^2\leq y^2}dx'dt'\biggl(\f{y^2-x'^2-t'^2}{y}\biggl)^{\Delta-2}\CO_\a(x+ix',t+t').\label{HKLLs}
\eal
We can check the equivalence of the two expressions (\ref{ours})(\ref{HKLLs}) by expanding in Taylor series of a primary operator in the HKLL expression;
\bal
\phi_\a^{\rm HKLL}(y,x,t)&=\f{\Delta-1}{\pi}\int_{x'^2+t'^2\leq y^2}dx'dt'\biggl(\f{y^2-x'^2-t'^2}{y}\biggl)^{\Delta-2}\sum_{m,n}\f{1}{m!n!}(x'^{+})^{m}(x'^{-})^{n}\pp_{+}^{m}\pp_{-}^{n}\CO_\a(x^+,x^-)\no
&=\sum_{k=0}^{\infty}y^{\Delta+2k}\f{\Gamma(\Delta)}{k!\Gamma(k+\Delta)}\pp_+^{k}\pp_-^k\CO_\a(x^+,x^-)\label{poi}
\eal
Since $\pp_+^{k}\pp_-^k=%(-1)^k {\bar{L}^P_{\pm}\equiv -i{\bar{L}_1+\bar{L}_3}; Ishibashi \r twisted ishibashi}
-(L^h_{-1})^{k}(\bar{L}^h_{-1})^{k}$, we can conclude
\ba
\phi_\a^{\rm HKLL}(y,x,t)|0\lb_{\rm CFT}=|\phi_\a(y,x,t)\lb.
\ea
\subsection{Representation in the hyperbolic basis (I\hspace{-.1em}I)}
We comment on the basis which diagonalizes a non-compact direction $J_1$.
\begin{alignat}{2}
L^R_0&=-J_1,\q L^R_{\pm 1}&=&-(J_0\pm J_2),\no
\bar{L}^R_0&=\bar{J}_1,\qq \bar{L}^R_{\pm 1}&=&\bar{J}_0 \mp \bar{J}_2
\end{alignat}
From (\ref{commuj}), they satisfy the following commutation relations
\ba
[L^R_0,L^R_{\pm 1}]=\mp L^R_{\pm 1},\q [L^R_{+1},L^R_{-1}]=2L^R_0.
\ea
Notice that $L^R_n$ are  anti-Hermitian operators.\\
The explicit relation with the elliptic basis is expressed as follows:
\begin{alignat}{2}
L^{R}_0&=i\f{L_1+L_{-1}}{2},
&\bar{L}^{R}_0&=-i\f{\bar{L}_1+\bar{L}_{-1}}{2} \no L^{R}_{\pm 1}&=iL_0\pm\f{L_1-L_{-1}}{2},&\bar{L}^{R}_{\pm 1}&=-i\bar{L}_0\pm\f{\bar{L}_1-\bar{L}_{-1}}{2},\no
L_0&=-i\f{L^{R}_1+L^{R}_{-1}}{2},&\bar{L}_0&=i\f{\bar{L}^{R}_1+\bar{L}^{R}_{-1}}{2}
\no L_{\pm 1}&=-iL^{R}_0\pm \f{L^{R}_1-L^{R}_{-1}}{2}
,&\bar{L}_{\pm 1}&=i\bar{L}^{R}_0\pm\f{\bar{L}^{R}_1-\bar{L}^{R}_{-1}}{2}.
\end{alignat}
These two bases are related by the following non-unitary transformation
\bal
U_e^RL_n(U_e^R)^{-1}&=L^R_n\no
U_e^R\bar{L}_n(U_e^R)^{-1}&=\bar{L}^R_n,\ U_e^R=e^{\f{i\pi}{4}(L_1-L_{-1}-\bar{L}_1+\bar{L}_{-1})}\no
U_R^eL^{R}_n(U_R^e)^{-1}&=L_n\no
U_R^e\bar{L}^{R}_n(U_R^e)^{-1}&=\bar{L}_n,\ U_e^R=e^{-\f{i\pi}{4}(L^{R}_1-L^{R}_{-1}-\bar{L}^{R}_1+\bar{L}^{R}_{-1})}.
\eal
\textbf{Rindler coordinate}\\
In the Rindler coordinate, $SL(2,\mathbb{R})$ generators can be written as
\bal
L^{r,l}_0&=-\pp_+\no
L^{r,l}_{\pm 1}&=-\f{\sqrt{r^2-R^2}}{2r}e^{\pm x^{+}}\biggl[ \f{2r^2-R^2}{r^2-R^2}\pp_+-\f{R^2}{r^2-R^2}\pp_-\mp r\pp_r \biggl]\no
\bar{L}^{r,l}_0&=-\pp_-\no
\bar{L}^{r,l}_{\pm 1}&=-\f{\sqrt{r^2-R^2}}{2r}e^{\pm x^{-}}\biggl[\f{2r^2-R^2}{r^2-R^2}\pp_--\f{R^2}{r^2-R^2}\pp_+\mp r\pp_r \biggl].
\eal
As we saw in section \ref{Rindler}, $L_{n}^{l}$ and $L_{n}^{r}$ are if combined, become globally defined $SL(2,\mathbb{R})$ generators in the Euclidean coordinate $\zeta=e^{\phi_{r}-i\tau_{r}}=e^{-\phi_{l}-i(\tau_{l}+\pi)},\bar{\zeta}=e^{\phi_{r}+i\tau_{t}}=e^{-\phi_{l}+i(\tau_{l}+\pi)}$ defined on the boundary of AdS. In particular, it is identical to the basis of $SL(2,\mathbb{R})$ diagonalizing $J_1$.
  \bal
L^{R}_{n}& \equiv
\int_{\mathbb{R}} d\zeta \zeta^{n+1}T(\zeta)=L^{r}_{n}-(-1)^{n}L^{l}_{-n}\no
\bar{L}^{R}_{n}& \equiv
\int_{\mathbb{R}} d\bar{\zeta}\bar{\zeta}^{n+1}\bar{T}(\bar{\zeta})=\bar{L}^{r}_{n}-(-1)^{n}\bar{L}^{l}_{-n}
\eal
The conjugates of $L_{n}$ are defined by $z\r \z$, thus $L^R_{n}$ are anti-Hermite operators
\ba
L_{n}^{R\dg}=-L^R_{n}.
\ea
The Rindler Hamiltonian can be written as $H^{\rm Rindler}=-i(L_0-\bar{L}_0)$ thus it is  diagonalized in this basis. Again the conditions for the bulk local states can be written in this basis as
\ba
(L^R_0-\bar{L}^R_0)|\phi_\a\lb=(L^R_{\pm 1}+\bar{L}^R_{\mp 1})|\phi_\a\lb=0.
\ea
This is just the same condition as in the elliptic basis. Actually the bulk local states are invariant under the conformal transformation between the elliptic basis and this hyperbolic basis
\ba
U^R_e|\phi_\a\lb=|\phi_\a\lb.
\ea
We can express the bulk local states in terms of this basis just the same as we did for $L^h_n$ in the previous subsection. 
\section{Comparison with mode expansions of bulk local operators}
The bulk local operator in the global AdS coordinate an be expanded by the creation/annihilation operators $a^{\dg}_{h,\bar{h}}$ as \cite{Kaplan}
\ba
\phi(r,t,\phi)=\sum_{w,k}\f{1}{\CN^{\CO}_{n,l}}[e^{-i\w t}Y_{l}(\phi)f_{h,\bar{h}}(r)a_{h,\bar{h}}+e^{i\w t}Y^{\dg}_{l}(\phi)f^{\dg}_{h,\bar{h}}(r)a^{\dg}_{h,\bar{h}}],\label{blo}
\ea
where $\w=\Delta+2n+|l|,Y_{l}(\phi)=e^{il\phi}$ and $2n+|l|=h+\bar{h}, l=h-\bar{h}$. $f_{h,\bar{h}}(r)$ is expressed as 
\bal
&f_{h,\bar{h}}(r)=\biggl(\f{r^{2}}{1+r^{2}}\biggl)^{\f{w}{2}}r^{-\Delta}{}_{2}F_{1}(\f{\Delta-l-\w-d+2}{2},\f{\Delta+l-\w}{2},1-\f{d}{2}+\Delta;-\f{1}{r^{2}})\no
&=\sin^{(2n+|l|)}\eta\cos^{\Delta}\eta{}_{2}F_{1}(-n-|l|-\f{d-2}{2},-n,1-\f{d}{2}+\Delta;-\f{1}{\tan^{2}\eta})\no
&=\sin^{|l|}\eta\cos^{\Delta}\eta{}_{2}F_{1}(\Delta+n+|l|,-n,1-\f{d}{2}+\Delta;\cos^{2}\eta)\no
&=\f{\Gamma(n+|l|+\f{d}{2})\Gamma(\f{d}{2}-n-\Delta)}{\Gamma(|l|+\f{d}{2})\Gamma(\f{d}{2}-\Delta)}\sin^{|l|}\eta\cos^{\Delta}\eta{}_{2}F_{1}(\Delta+n+|l|,-n,|l|+\f{d}{2};\sin^{2}\eta)
+\no&\f{\Gamma(1-\f{d}{2}+\Delta)\Gamma(|l|+\f{d}{2}-1)}{\Gamma(\Delta+n+|l|)\Gamma(-n)}
\sin^{2-d}\eta\cos^{\Delta}\eta{}_{2}F_{1}(\f{2-d}{2}-n-|l|,\f{2-d}{2}+\Delta+n,1+\f{2-d}{2}-|l|;\sin^{2}\eta)\no
&=\f{\Gamma(n+|l|+\f{d}{2})\Gamma(\f{d}{2}-n-\Delta)}{\Gamma(|l|+\f{d}{2})\Gamma(\f{d}{2}-\Delta)}\sin^{|l|}\eta\cos^{\Delta}\eta{}_{2}F_{1}(\Delta+n+|l|,-n,|l|+\f{d}{2};\sin^{2}\eta)\no
&=\f{\Gamma(n+|l|+\f{d}{2})\Gamma(\f{d}{2}-n-\Delta)}{\Gamma(|l|+\f{d}{2})\Gamma(\f{d}{2}-\Delta)}\sinh^{|l|}\rho\cosh^{\Delta+2\Delta+|l|}\rho{}_{2}F_{1}(\Delta+n+|l|,n+|l|+\f{d}{2},|l|+\f{d}{2};-\sinh^{2}\rho),\nonumber
\eal
where $r=\tan\eta=\sinh\rho$ and we used the following formula
\bal
{}_{2}F_{1}(a,b,c;z)&=(1-z)^{-b}{}_{2}F_{1}(c-a,b,c;z/(z-1))\no
{}_{2}F_{1}(a,b,c;z)&=\f{\Gamma(c)\Gamma(c-a-b)}{\Gamma(c-a)\Gamma(c-b)}{}_{2}F_{1}(a,b,a+b+1-c:1-z)\no&+\f{\Gamma(c)\Gamma(a+b-c)}{\Gamma(a)\Gamma(b)}(1-z)^{c-a-b}{}_{2}F_{1}(c-a,c-b,-a-b+1+c:1-z),\nonumber
\eal
and used the fact that $1/\Gamma(-n)=0$. The normalization factor is evaluated as
\ba
\f{1}{\CN^{\CO}_{n,l}}\propto
(-1)^{n}\f{\Gamma(l+\f{d}{2})\Gamma(\f{d}{2}-\Delta)}{\Gamma(n+l+\f{d}{2})\Gamma(\f{d}{2}-n-\Delta)}\biggl(\f{n!\Gamma^{2}(l+\f{d}{2})\Gamma(\Delta+n-\f{d-2}{2})}{\Gamma(n+l+\f{d}{2})\Gamma(\Delta+n+l)}\biggl)^{-1/2}.%\no
%\propto\sqrt{\f{\Gamma(\Delta+n+l)\Gamma(\Delta+n-\f{d-2}{2})\Gamma(\f{d}{2})}{n!\Gamma(\Delta)\Gamma(\Delta-\f{d-2}{2})\Gamma(\f{d}{2}+n+l)}}
\ea
From the extrapolation dictionary of AdS/CFT, we obtain
\ba
\lim_{r\r\infty}r^{\Delta}\phi(r,t,\phi)=\sum_{w,k}\f{1}{\CN^{\CO}_{n,l}}[e^{-i\w t}Y_{l}(\phi)a_{h,\bar{h}}+e^{i\w t}Y^{\dg}_{l}(\phi)a^{\dg}_{h,\bar{h}}]=\CO(t,\phi).
\ea
Thus the dual primary operator can be expanded by $a^{\dg}_{h,\bar{h}}$ as follows 
\ba
\CO(t,\phi)=\sum_{w,k}\f{1}{\CN^{\CO}_{n,l}}[e^{-i\w t}Y_{l}(\phi)a_{h,\bar{h}}+e^{i\w t}Y^{\dg}_{l}(\phi)a^{\dg}_{h,\bar{h}}],
\ea
We move to the Euclidean plane by the conformal transformation $z=e^{\tau+i\phi},\z=e^{\tau-i\phi}$, where the primary operators can be expressed as
\bal
\CO(z,\z)&=\sum_{w,k}\f{1}{\CN^{\CO}_{n,l}}[(z\z)^{-(\Delta+n+|l|/2)}\biggl(\f{z}{\z}\biggl)^{\f{l}{2}}a_{n,l}+(z\z)^{n+|l|/2}\biggl(\f{z}{\z}\biggl)^{-\f{l}{2}}a^{\dg}_{n,l}]\no
&=\sum_{h,\bar{h}=0}^{\infty}\f{1}{\CN^{\CO}_{n,l}}[(z\z)^{-\Delta}z^{-h}\z^{-\bar{h}}a_{h,\bar{h}}+z^{h}\z^{\bar{h}}a^{\dg}_{h,\bar{h}}].
\eal
Primary state can be regarded as the lowest energy one particle state in AdS
\ba
\CO(0)|0\lb=a^{\dg}_{0,0}|0\lb
\ea
whose mass is $m^2=\Delta(\Delta-2)$.\\
The descendant states are dual to the higher energy one particle states
\ba
(P_{z})^{h}(P_{\z})^{\bar{h}}\CO(0)|0\lb=(\pp)^{h}(\db)^{\bar{h}}\CO(0)|0\lb=\f{h!\bar{h}!}{\CN^{\CO}_{h,\bar{h}}}a^{\dg}_{h,\bar{h}}|0\lb.
\ea
\\
The wave function for the descendant state $|h,\bar{h}\lb=a^{\dg}_{h,\bar{h}}|0\lb$ is calculated as
\bal
\psi_{n,l}(r,t,\phi)&=\la \phi(r,t,\phi)|h,\bar{h}\lb\no&=\la \phi(r,t,\phi)|\f{\CN^{\CO}_{n,l}}{h!\bar{h}!}(P_{z})^{h}(P_{\z})^{\bar{h}}|\CO\lb\no&=\f{1}{\CN^{\CO}_{n,l}}e^{-i\w t}Y_{l}(\phi)f_{h,\bar{h}}(r)\no&=\f{1}{\CN_{\Delta nl}}e^{-i\w t}Y_{l}(\phi)\sin^{|l|}\rho\cos^{\Delta}\rho{}_{2}F_{1}(\Delta+n+|l|,-n,|l|+\f{d}{2};\sin^{2}\eta),
\eal
where
\ba
\CN_{\Delta nl}=(-1)^{n}\sqrt{\f{n!\Gamma^{2}(l+\f{d}{2})\Gamma(\Delta+n-\f{d-2}{2})}{\Gamma(n+l+\f{d}{2})\Gamma(\Delta+n+l)}}.
\ea
One can check that the equivalence of the expression of the bulk local operators (\ref{blo}) and the twisted Ishibashi states $|\phi\lb$ as follows. We can evaluate the function $f_{h,\bar{h}}(r)$ at the center of the AdS:
\bal
f_{h=\bar{h}}(r=0)&=(-1)^{n}\f{\Gamma(n+1)\Gamma(\Delta)}{\Gamma(n+\Delta)}\no
f_{h\neq\bar{h}}(r=0)&=0.
\eal
From the expression (\ref{blo}), the bulk local state $\phi(0,0,0)|0\lb$ can be expressed as
\bal
\phi(0,0,0)|0\lb&=\sum_{n}(-1)^{n}\f{\Gamma(n+1)\Gamma(\Delta)}{\Gamma(n+\Delta)}\f{1}{\CN^{\CO}_{n,l}}a^{\dg}_{n=h=\bar{h}}|0\lb\no
&=\sum_{n}(-1)^{n}\f{\Gamma(\Delta)}{\Gamma(n+1)\Gamma(n+\Delta)}(P_{z})^{n}(P_{\z})^{n}\CO(0)|0\lb.
\eal
Thus the expression (\ref{blo}) can be reduced to the twisted Ishibashi state.

We can express creation/annihilation operators as integrals of primary operators:
\ba
a_{\w,l}=\CN^{\CO}_{n,l}\int\f{dt'd\phi'}{(2\pi)^{2}}e^{iwt'-il\phi'}\CO(t',\phi').
\ea
Thus we can express the bulk local operators as integral of primary operators
\bal
\phi(r,t,\phi)&=\int dt'd\phi'\biggl(\f{1}{(2\pi)^{2}}\sum_{\w,l}e^{-i\w(t-t')+il(\phi-\phi')}f_{\w,l}(r)\biggl)\CO(t',\phi')\no&=
\int dt'd\phi'K(r,t,\phi;t',\phi')\CO(t',\phi').
\eal
where we defined
\ba
K(r,t,\phi;t',\phi')=\f{1}{(2\pi)^{2}}\sum_{\w,l}e^{-i\w(t-t')+il(\phi-\phi')}f_{\w,l}(r).
\ea
This is expression known as the HKLL prescription. We can check the equivalence between the expression by the HKLL construction and the twisted Ishibashi state more directly \cite{GMT}.

\end{document}